\documentclass[12pt]{iopjournal}

\usepackage{amsmath, amsfonts, multirow, amssymb}
\usepackage{lmodern}
\usepackage{float}
\usepackage{ragged2e}   
\usepackage{microtype}  
\usepackage{makecell} 

\DeclareMathOperator{\sgn}{sgn}

\microtypesetup{
  activate = true,
  expansion = true,
  protrusion = true
}
\usepackage[numbers,square]{natbib}
\AtBeginDocument{\justifying}

\usepackage{etoolbox}
\AtBeginEnvironment{abstract}{\justifying}

\AtBeginDocument{%
  \pagestyle{plain}
  \thispagestyle{plain}
}

\usepackage{sectsty}
\sectionfont{\Large\bfseries}         
\subsectionfont{\large\bfseries}      
\subsubsectionfont{\normalsize\bfseries}

\usepackage{caption}
\begin{document}


\title{\LARGE Extremal and global-time statistics of sluggish random walks with stochastic resetting}

\author{Theophilus Dwamena Frimpong$^{1,*}$}

\affil{$^1$The University of Texas Rio Grande Valley, 1201 W. University Dr., Edinburg, TX 78539, USA}



\begin{abstract}
We study a sluggish random walk, whose diffusivity decreases as a power law with distance from the origin, under Poissonian resetting to the origin. Using a mapping to a signed Bessel process and renewal theory, we obtain exact results for the maximum trajectory value in a finite window, the time at which it occurs, and the time of the last visit to the origin. Resetting changes the growth of the typical maximum from a power law in time to a power of its logarithm. The centered maximum approaches a Gumbel law, whose width shrinks in physical space for any positive sluggishness exponent, unlike the Brownian case. In the Bessel coordinate, the probability that the maximum exceeds a value growing in proportion to the observation window length decays exponentially, with a rate that has a second-order dynamical transition and is independent of sluggishness. The time of the maximum is uniform in the bulk, with boundary layers that shift its mean above the midpoint logarithmically with window length. The time since the last visit to the origin follows a Gamma law with a sluggishness-dependent shape. Reflection symmetry makes the last-visit time uncorrelated with the occupation time of the positive half-line.
\end{abstract}

\section{Introduction}
\label{sec:intro}

A stochastic trajectory observed over a finite time window contains information that its final position does not. For a one-dimensional trajectory $x(\tau)$ that starts at the origin $x_0$ and is observed for a time $T$, four path functionals are of particular interest: the maximum $M_T$ (largest value of $x(\tau)$), the time $t_M$ at which it is reached, the last-passage time $t_l$ through the origin, and the occupation time $t_+$ spent on the positive half-line. The maximum $M_T$ of a trajectory and the time $t_M$ at which it is reached belong to extreme-value statistics, which are well understood for independent variables but difficult to apply to the correlated values along a trajectory~\cite{majumdar2005brownian,majumdar2020extreme,majumdar2024statistics,fortin2015applications}. For Brownian motion, the occupation time $t_+$, the last-passage time $t_l$ and the time of the maximum $t_M$ all have the same distribution,
\begin{equation}
P(t_c\,|\,T)=\frac{1}{\pi\sqrt{t_c\,(T-t_c)}},\qquad 0<t_c<T,\quad c\in\{+,\,l,\,M\},
\label{eq:arcsine}
\end{equation}
which is L\'evy's arcsine law~\cite{levy1940certains,feller1991introduction,morters2010brownian}. The density of this distribution is U-shaped, so the mean value $T/2$ is the least likely outcome. Arcsine laws have been generalized to fractional Brownian motion~\cite{sadhu2018generalized}, run-and-tumble particles~\cite{singh2019generalised}, continuous-time random walks~\cite{bel2005weak} and power-law heterogeneous diffusion~\cite{singh2022extreme}. In general the distributions of $t_+$, $t_l$ and $t_M$ no longer coincide~\cite{sadhu2018generalized,del2025generalized}, and even for Brownian motion their joint statistics are nontrivial~\cite{hartmann2025exact}.

Stochastic resetting is a mechanism in which a process is interrupted at random times and restarted from a prescribed state~\cite{evans2011diffusion,evans2020stochastic,gupta2022brief}. We consider Poissonian resetting at a constant rate $r$, so the waiting times between resets are independent and exponentially distributed with mean $1/r$. A diffusing particle then relaxes to a nonequilibrium stationary state, and its mean first-passage time to a target is finite and minimal at an optimal rate~\cite{evans2011diffusion,evans2020stochastic,reuveni2016optimal}. 

Resetting reshapes the extremal and global-time statistics of a trajectory. For Brownian motion with resetting to the origin, the maximum grows only logarithmically in time and follows a Gumbel law~\cite{singh2021extremal,majumdar2021mean}, which is one of the three classical limit laws of extreme-value statistics~\cite{fisher1928limiting,gnedenko1943distribution,gumbel2004statistics}. In the far tail, where the maximum grows linearly in time, its large-deviation function has a second-order dynamical transition~\cite{hartmann2024distribution}. The time of the maximum becomes uniformly distributed over the observation window~\cite{singh2021extremal}. This follows from the renewal structure of resetting. After each reset, the process starts afresh, so the trajectory is a sequence of statistically independent pieces, each lasting from one reset to the next. We call each such piece an inter-reset interval, or simply an \emph{interval}. At long times, the global maximum is equally likely to lie in any completed interval~\cite{singh2021extremal}. The uniformity fails near the two ends of the observation window, for two different reasons. At the right end, the interval that is running when the window closes is cut off at time $T$, so its duration is not exponentially distributed like that of a completed interval. At the left end, a walker that starts at the origin needs a finite time to climb to a high level, so an early maximum is suppressed. For Brownian motion with resetting that starts in its stationary state rather than at the reset point, the time of the maximum is also uniform in the bulk, and its two edge regimes have been computed~\cite{mori2022time}. For resetting to the origin, the occupation time remains symmetric about $T/2$ at any $r$, the last-passage time is biased toward late times~\cite{tazbierski2025last}, and the pairwise correlations among the three times depend nontrivially on $r$~\cite{wang2026pairwise}.

In heterogeneous media the diffusivity depends on position, and power-law profiles produce anomalous diffusion and ergodicity breaking~\cite{fa2003power,cherstvy2013anomalous,leibovich2019infinite,stella2023universal}. The sluggish random walk is an exactly solvable model of this kind. It is a lattice walk whose hopping probability decreases algebraically with the distance from the origin. In the continuum limit it becomes a diffusion with $D(x)=|x|^{-\alpha}$, $\alpha\ge0$, whose typical displacement grows subdiffusively as $t^{\mu}$ with $\mu=1/(\alpha+2)$, and whose position distribution is bimodal at late times for $\alpha>0$~\cite{zodage2023sluggish,boyer2026ultraslow}. For this walk, the distributions of $t_+$, $t_l$ and $t_M$ are no longer identical and depend on $\alpha$~\cite{del2025generalized}. Resetting in heterogeneous media has been analyzed for several diffusivity profiles and stochastic-calculus conventions~\cite{ray2020space,lenzi2022transient,menon2024random,ray2022expediting,delvecchio2026proxitaxis}. For the sluggish walk reset to the origin, the stationary distribution, first-passage statistics and occupation time have been characterized~\cite{frimpong2026}, and resetting to previously visited positions slows the growth of the typical position to $[\ln(rt)]^{1/(\alpha+2)}$~\cite{boyer2026ultraslow}.

To our knowledge, the maximum, the time of the maximum and the last-passage time of a sluggish walk with Poissonian resetting to the origin have not been determined, although the reset-free sluggish walk~\cite{del2025generalized} and Brownian motion with resetting~\cite{singh2021extremal,tazbierski2025last} are known. The Brownian results cannot simply be carried over. Brownian motion is invariant under translations, so with resetting the statistics of the maximum depend on the level only through its distance from the reset point. For $\alpha>0$ the diffusivity $D(x)=|x|^{-\alpha}$ breaks this invariance, and the reset point is also the point where the diffusivity is largest. Shifting the origin therefore does not reduce the problem to the Brownian one, and the question is how resetting, which returns the walker to the region of high diffusivity and confines it near the origin, competes with the decrease of the diffusivity at large distances.

Here we address this question for Poissonian resetting to the origin. A monotonic transformation maps the heterogeneous diffusion onto a symmetric two-sided Bessel process of dimension $d=2(1-\mu)<2$. This representation, together with renewal theory, yields exact Laplace-domain results for the maximum, the time at which it occurs, and the last-passage time. The occupation time of this process was characterized in Ref.~\cite{frimpong2026}; here it enters only through a reflection-symmetry relation with the last-passage time.

The paper is organized as follows. Section~\ref{sec:building_blocks} defines the sluggish random walk with Poissonian resetting model, maps it onto a signed Bessel process, and derives the renewal relations under resetting. Section~\ref{sec:maximum} derives the exact distribution of the maximum for all $\alpha$, its Gumbel limit and its far-tail transition. Section~\ref{sec:tmax} treats the time of the maximum, including its uniform bulk and boundary layers, and Sec.~\ref{sec:lastpassage} the last-passage time and its symmetry relation with the occupation time. Section~\ref{sec:conclusions} summarizes our findings. Technical derivations are collected in Appendix~A and the numerical implementation in Appendix~B.

\section{Model and analytical framework}
\label{sec:building_blocks}

\subsection{Sluggish random walk with resetting and its observables}
\label{subsec:model}

We consider a walker on a one-dimensional lattice of spacing $b$. In each time step $\Delta t$ the walker is reset to the origin with probability $p_{\rm reset}=r\Delta t+o(\Delta t)$. Otherwise it hops to each neighboring site $x\pm b$ with probability $1/[(|x|/b)^{\alpha}+2]$ and stays at $x$ with the remaining probability~\cite{zodage2023sluggish}. The exponent $\alpha\ge0$ controls the sluggishness, because the hopping probability decreases with the distance from the origin. The continuum limit is taken with $b$ and $\Delta t$ sent to zero at fixed $b^{\alpha+2}/\Delta t=1$.\footnote{For $\alpha=0$ the constant $2$ in the denominator is not negligible, and one takes $b^{2}/\Delta t=3$ instead, which again gives a unit diffusion constant.} The probability density $P(x,t)$ of the position then obeys~\cite{zodage2023sluggish,del2025generalized, frimpong2026}
\begin{equation}
\frac{\partial P}{\partial t}=\frac{\partial^2}{\partial x^2}\Bigl[|x|^{-\alpha}P\Bigr]-rP+r\,\delta(x),
\label{eq:FP}
\end{equation}
where $r$ is the resetting rate and $\delta$ is the Dirac delta function. Between resets the position follows the It\^o equation $dx=\sqrt{2|x|^{-\alpha}}\,dW_t$, where $W_t$ is a standard Wiener process~\cite{zodage2023sluggish}. For $\alpha>0$ the extension of this equation through the origin (Sec.~\ref{subsec:bessel}) is part of the model definition. Other stochastic-calculus conventions, or other extensions through the origin, define different processes, and all results below refer to the It\^o form of Eq.~\eqref{eq:FP} with the symmetric, non-sticky extension.

The walker starts at the origin and is reset to the origin. After each reset it therefore starts afresh from the same state, and the trajectory is a sequence of independent inter-reset intervals, the last of which is truncated by the end of the observation window. Throughout we write
\begin{equation}
\mu=\frac{1}{\alpha+2}\in\Bigl(0,\tfrac12\Bigr].
\label{eq:mu}
\end{equation}

For a trajectory of duration $T$ we study the maximum and the time at which it is reached,
\begin{equation}
M_T=\max_{0\le\tau\le T}x(\tau),\qquad t_M=\operatorname*{argmax}_{0\le\tau\le T}x(\tau),
\label{eq:max_def}
\end{equation}
together with the last-passage time through the origin and the occupation time of the positive half-line,
\begin{equation}
t_l=\sup\{t\le T:\ x(t)=0\},\qquad t_+=\int_0^T\Theta\bigl(x(\tau)\bigr)\,d\tau ,
\label{eq:tl_def}
\end{equation}
where $\Theta$ is the Heaviside step function.\footnote{After a reset the walker crosses the origin infinitely often, so counting or not counting the reset instants as passages gives the same $t_l$ with probability one.} Equation~\eqref{eq:FP} contains a single length scale $\ell_r=r^{-\mu}$ and a single time scale $1/r$. Every time distribution therefore has the form $p(t|T)=T^{-1}\mathcal F_\alpha(t/T,rT)$, where $\mathcal F_\alpha$ is a dimensionless function of the scaled time $t/T$ and of the product $rT$, the mean number of resets in the window. The maximum can be written as $M_T=r^{-\mu}\mathcal M_\alpha(rT)$, where $\mathcal M_\alpha$ is a dimensionless random variable whose law depends only on $rT$.

\subsection{Mapping to a signed Bessel process}
\label{subsec:bessel}

The position-dependent diffusivity complicates the equations for all observables above. It can be removed by the change of variable
\begin{equation}
y(x)=\sgn(x)\,\frac{2}{\alpha+2}\,|x|^{(\alpha+2)/2}=2\mu\,\sgn(x)\,|x|^{1/(2\mu)}.
\label{eq:ymap}
\end{equation}
For $x>0$ one has $dy/dx=x^{\alpha/2}$, and the chain rule gives
\begin{equation}
|x|^{-\alpha}\partial_x^2=\partial_y^2+\frac{\alpha}{2}\,x^{-(\alpha+2)/2}\,\partial_y=\partial_y^2+\frac{1-2\mu}{y}\,\partial_y ,
\label{eq:generator_y}
\end{equation}
where we used $x^{-(\alpha+2)/2}=2\mu/y$ and $\alpha\mu=1-2\mu$. For $x<0$ the same signed-coordinate generator holds with $y<0$. Equivalently, the radial coordinate $|y|$ follows the same Bessel diffusion. The right-hand side of Eq.~\eqref{eq:generator_y} is the generator of a diffusion with unit diffusion constant and a repulsive drift $(1-2\mu)/y$ (an It\^o correction).

During each excursion away from the origin, $|y|$ is therefore a Bessel process of dimension
\begin{equation}
d=2(1-\mu)=\frac{2(\alpha+1)}{\alpha+2}\in[1,2)
\label{eq:dimension}
\end{equation}
with diffusion constant one~\cite{revuz2013continuous,borodin2012handbook}, that is, a diffusion with generator $\partial_y^2+[(d-1)/y]\,\partial_y$. For $\alpha=0$ one has $d=1$ and $y=x$. Since $d<2$, the process reaches the origin in finite time and the walk is recurrent. For a walker started away from the origin, the probability of not having reached the origin by time $t$ decays as $t^{-\mu}$, so $\mu=1-d/2$ is the persistence exponent~\cite{bray2013persistence,zodage2023sluggish,del2025generalized}. Bessel processes also appear in other extreme-value problems~\cite{schehr2010extreme}. For $\alpha>0$ the diffusivity diverges at the origin, so the It\^o equation does not fix the process there, and we define it as the time-changed Brownian motion $x(t)=B(\tau(t))$, where $B$ is a Brownian motion with unit diffusion constant, $\tau$ is the operational time, and $dt=|B|^{\alpha}d\tau$~\cite{revuz2013continuous,borodin2012handbook}. In the variable $y$ this is the symmetric two-sided Bessel diffusion, in which each excursion from the origin is a Bessel process of dimension $d$ and carries the sign $\pm$ with equal probability (Appendix~\ref{app:bessel}). Because the map~\eqref{eq:ymap} is increasing and preserves the sign of $x$, the times $t_+$, $t_l$ and $t_M$ are identical for the sluggish walk and for its Bessel image. Only the maximum must be mapped back. If $Y=y(M)$ denotes the maximum in the Bessel variable, then $M=\mathcal X(Y)$ with
\begin{equation}
\mathcal X(Y)=\Bigl(\frac{Y}{2\mu}\Bigr)^{2\mu},\qquad Y>0.
\label{eq:xinv}
\end{equation}

\subsection{Solutions of the backward equation}
\label{subsec:backward}

All observables below follow from the Laplace-space backward equation $|x|^{-\alpha}\psi''(x)=s\,\psi(x)$ for a function $\psi$ of the starting position $x$, where $s>0$ is the Laplace variable conjugate to time. For $x>0$ the scaled Bessel coordinate $z=\sqrt s\,y(x)$ turns it into $\psi_{zz}+(1-2\mu)\psi_z/z-\psi=0$, and the substitution $\psi=z^{\mu}f(z)$ gives the modified Bessel equation of order $\mu$~\cite{olver2010nist}. The solution that satisfies the matching condition of the symmetric extension at the origin, namely continuity of $\psi$ and $\psi'$ at $x=0$, and that is bounded as $x\to-\infty$ is (Appendix~\ref{app:bessel})
\begin{equation}
h_s(x)=\begin{cases} g_\mu\bigl(\sqrt s\,y(x)\bigr), & x\ge0,\\[10pt] \dfrac{2\sin\pi\mu}{\pi}\,z^{\mu}K_\mu(z),\quad z=\sqrt s\,|y(x)|, & x<0,\end{cases}
\label{eq:hs}
\end{equation}
where $K_\mu$ is the modified Bessel function of the second kind and
\begin{equation}
g_\mu(z)=z^{\mu}\bigl[I_{\mu}(z)+I_{-\mu}(z)\bigr],
\label{eq:gmu}
\end{equation}
with $I_{\pm\mu}$ the modified Bessel functions of the first kind. The function $g_\mu$ also appears in the time of the maximum of the free sluggish walk~\cite{del2025generalized}. It satisfies $g_\mu(0)=c_\mu$ with
\begin{equation}
c_\mu=\frac{2^{\mu}}{\Gamma(1-\mu)} ,
\label{eq:cmu}
\end{equation}
and $g_\mu''+(1-2\mu)g_\mu'/z-g_\mu=0$. Its large-argument behavior is $g_\mu(z)\simeq\sqrt{2/\pi}\,z^{\mu-1/2}e^{z}\bigl[1-(4\mu^2-1)/(8z)+\dots\bigr]$~\cite{olver2010nist}, and its derivative is $g_\mu'(z)=z^{\mu}\bigl[I_{\mu-1}(z)+I_{1-\mu}(z)\bigr]$. We write
\begin{equation}
\eta(z)=\frac{g_\mu'(z)}{g_\mu(z)},
\end{equation}
which tends to one as $z\to\infty$. For $\alpha=0$ one has $g_{1/2}(z)=\sqrt{2/\pi}\,e^{z}$ and $\eta=1$.

\subsection{Survival probability below an absorbing level}
\label{subsec:survival_level}

Both the maximum level $M_T$ and its time $t_M$ are controlled by the probability that the walker avoids a given level. Let $Q_0(x_0,t\,|\,M)$ be the probability that a free walker (without resetting) started at $x_0<M$ has not reached the level $M>0$ up to time $t$. It obeys the backward equation $\partial_tQ_0=|x_0|^{-\alpha}\partial_{x_0}^2Q_0$ with $Q_0(M,t\,|\,M)=0$ and $Q_0(x_0,0\,|\,M)=1$. Its Laplace transform $\tilde Q_0(x_0,s\,|\,M)=\int_0^\infty e^{-st}Q_0\,dt$ satisfies
\begin{equation}
|x_0|^{-\alpha}\,\partial_{x_0}^2\tilde Q_0-s\,\tilde Q_0=-1,\qquad \tilde Q_0(M,s\,|\,M)=0,
\label{eq:survival_ode}
\end{equation}
and stays bounded as $x_0\to-\infty$. A particular solution is $1/s$, and the only solution of the homogeneous equation that is bounded at $-\infty$ is $h_s$. Hence $\tilde Q_0=1/s+B\,h_s(x_0)$, and the absorbing condition at $M$ fixes the constant $B$, giving
\begin{equation}
\tilde Q_0(x_0,s\,|\,M)=\frac1s\left[1-\frac{h_s(x_0)}{g_\mu\bigl(\sqrt s\,Y\bigr)}\right],\qquad Y=y(M)=2\mu M^{1/(2\mu)} .
\label{eq:Q0_general}
\end{equation}
For a walker started at the origin, $h_s(0)=c_\mu$ and
\begin{equation}
\tilde Q_0(M,s)\equiv\tilde Q_0(0,s\,|\,M)=\frac{1}{s}\Bigl[1-G_\mu(M,s)\Bigr],\qquad G_\mu(M,s)=\frac{c_\mu}{g_\mu\bigl(\sqrt s\,Y\bigr)} .
\label{eq:Q0_origin}
\end{equation}
Since $Q_0(0,t\,|\,M)=\mathrm{Prob}[M_t<M]$ for the free walker, Eq.~\eqref{eq:Q0_origin} is the Laplace transform of the distribution function of its maximum. The function $G_\mu(M,s)$ has a direct meaning. It is the probability that a free walker started at the origin reaches the level $M$ before an independent exponential clock of rate $s$ rings. It decreases from $G_\mu=1$ at $M=0$ to zero as $M\to\infty$. Equation~\eqref{eq:Q0_origin} reproduces known cases. For $\alpha=0$ it gives the Brownian result $\tilde Q_0=[1-e^{-\sqrt s M}]/s$~\cite{redner2001guide}. For $\alpha=1$, that is $\mu=1/3$, the function $g_\mu$ is proportional to the Airy function $\mathrm{Bi}$ of the second kind, through $\mathrm{Bi}(w)=\sqrt{w/3}\,[I_{-1/3}+I_{1/3}](\tfrac23w^{3/2})$~\cite{olver2010nist}, and Eq.~\eqref{eq:Q0_origin} reduces to the Airy-function expression of Ref.~\cite{zodage2023sluggish}.

By dimensional analysis the free maximum has the scaling form $\mathrm{Prob}[M_t<M]=\mathcal Q_0(Y/\sqrt t)$, so $M_t$ is typically of order $t^{\mu}$. The small-$s$ and large-$s$ expansions of Eq.~\eqref{eq:Q0_origin} give the two limits
\begin{align}
\mathrm{Prob}[M_t<M]&\simeq\frac{1}{\Gamma(1+\mu)}\Bigl(\frac{Y^2}{4t}\Bigr)^{\mu}\ \ (Y^2\ll t),\notag\\[4pt]
\mathrm{Prob}[M_t\ge M]&\simeq\frac{1}{\Gamma(1-\mu)}\Bigl(\frac{4t}{Y^2}\Bigr)^{\mu}e^{-Y^2/4t}\ \ (Y^2\gg t).
\label{eq:Q0_limits}    
\end{align}
The first limit $(Y^2\ll t)$ is the persistence law, the decay $t^{-\mu}$ of the probability to stay below a small level, with an explicit amplitude. Because $Y^{2\mu}=(2\mu)^{2\mu}M$, it also shows that the density of the free maximum is finite at $M=0$. The second limit $(Y^2\gg t)$ is a Gaussian far-right tail in the Bessel variable, which for $\alpha>0$ decays faster than a Gaussian in $M$ because $Y^2\propto M^{\alpha+2}$. For $\alpha=0$ the two limits reduce to $M/\sqrt{\pi t}$ and to the large-argument form of $\mathrm{erfc}[M/(2\sqrt t)]$, where $\mathrm{erfc}$ is the complementary error function, in agreement with the Brownian result $\mathrm{Prob}[M_t<M]=\mathrm{erf}[M/(2\sqrt t)]$~\cite{redner2001guide,morters2010brownian}.

\subsection{Renewal relations under resetting}
\label{subsec:renewal}

The free first-passage density to a level $M$ is $F_0=-\partial_tQ_0$, since $Q_0$ is the probability that the level has not yet been reached. From Eq.~\eqref{eq:Q0_general},
\begin{equation}
\tilde F_0(x_0,s\,|\,M)=\frac{h_s(x_0)}{g_\mu(\sqrt s\,Y)},\qquad
\tilde F_0(M,s)\equiv\tilde F_0(0,s\,|\,M)=G_\mu(M,s).
\label{eq:F0}
\end{equation}
Its small-$s$ expansion is nonanalytic, $1-\tilde F_0\propto s^{\mu}$, so the mean first-passage time to $M$ is infinite without resetting. Resetting at rate $r$ renews the process after every reset~\cite{evans2011diffusion,evans2020stochastic}. Conditioning on the time $\tau$ of the first reset, the walker either survives below $M$ for the whole window without a reset, or it survives the first interval of duration $\tau$ and then restarts from the origin. For a walker started at the reset point, the survival probability of the reset process is
\begin{align}
Q_r(M,t)&=e^{-rt}Q_0(M,t)+r\int_0^t d\tau\,e^{-r\tau}Q_0(M,\tau)Q_r(M,t-\tau),\notag\\[4pt]
\tilde Q_r(M,s)&=\frac{\tilde Q_0(M,s+r)}{1-r\tilde Q_0(M,s+r)} .
\label{eq:renewal_a}
\end{align}
For a walker started at $x_0\ne0$ only the first interval starts away from the reset point, so
\begin{equation}
\tilde Q_r(x_0,s\,|\,M)=\tilde Q_0(x_0,s+r\,|\,M)\bigl[1+r\tilde Q_r(M,s)\bigr].
\label{eq:renewal_b_LT}
\end{equation}

\section{Distribution of the maximum}
\label{sec:maximum}

\subsection{Exact distribution}
\label{subsec:max_exact}

We first determine the distribution of the maximum. The event $M_T<M$ is the event that the walker has not reached the level $M$ up to time $T$, so the distribution function of $M_T$ is the survival probability of the reset process below an absorbing level at $M$. Inserting Eq.~\eqref{eq:Q0_origin} into the renewal relation~\eqref{eq:renewal_a} gives
\begin{equation}
\int_0^\infty dT\,e^{-sT}\,\mathrm{Prob}[M_T<M]=\tilde Q_r(M,s)=\frac{1-G_\mu(M,s+r)}{s+r\,G_\mu(M,s+r)} .
\label{eq:Qr}
\end{equation}
The probability density of the maximum is $p_r(M,T)=\partial_M\,\mathrm{Prob}[M_T<M]$. Equation~\eqref{eq:Qr} is exact for all $\alpha\ge0$, all $r$ and all $T$. For $\alpha=0$ it reduces to the Brownian result with $G_{1/2}=e^{-\sqrt{s+r}\,M}$~\cite{evans2011diffusion,singh2021extremal,debicki2026distributional}, and for $r=0$ it reduces to Eq.~\eqref{eq:Q0_origin}. Setting $s=0$ in Eq.~\eqref{eq:Qr} gives the mean first-passage time to the level $M$,
\begin{equation}
\langle T_M\rangle_r=\tilde Q_r(M,0)=\frac1r\left[\frac{g_\mu(\sqrt r\,Y)}{c_\mu}-1\right].
\label{eq:mfpt}
\end{equation}
It is finite for every $r>0$, whereas it diverges without resetting (Sec.~\ref{subsec:renewal}), and it reduces to the Brownian result at $\alpha=0$~\cite{evans2011diffusion}.

\subsection{Finite-time form and scaling}
\label{subsec:max_finite}

Equation~\eqref{eq:Qr} depends on the Laplace variable $s$, the resetting rate $r$ and the level $M$ only through the ratio $s/r$ and the rescaled level $\sqrt r\,Y$, up to an overall factor $1/r$, so that
\begin{equation}
\mathrm{Prob}[M_T<M]=\mathcal Q_\alpha(\sqrt r\,Y,rT).
\label{eq:Qr_scaling}
\end{equation}
Because $\sqrt r\,Y=2\mu\,(r^{\mu}M)^{1/(2\mu)}$, the distribution function at fixed $rT$ depends on the rate and the level only through $r^{\mu}M$, that is, through the level measured in units of $\ell_r$. For $rT\ll1$ resetting is rare and the free law is recovered, with $M_T$ of order $T^{\mu}$. For $rT\gg1$ the walker is reset many times and the maximum grows only as a power $2\mu\le1$ of $\ln(rT)$, which is logarithmic growth for $\alpha=0$, as shown in Sec.~\ref{subsec:max_gumbel}. The density of the maximum remains finite at $M=0$ for every $rT$. Only trajectories without a reset can stay below an arbitrarily small level, because each further interval costs another factor of the level, and Eq.~\eqref{eq:Q0_limits} gives
\begin{equation}
p_r(0,T)=e^{-rT}\,\frac{(2\mu)^{2\mu}}{(4T)^{\mu}\,\Gamma(1+\mu)} .
\label{eq:p0}
\end{equation}

\begin{figure}[h]
\centering
\includegraphics[width=\linewidth]{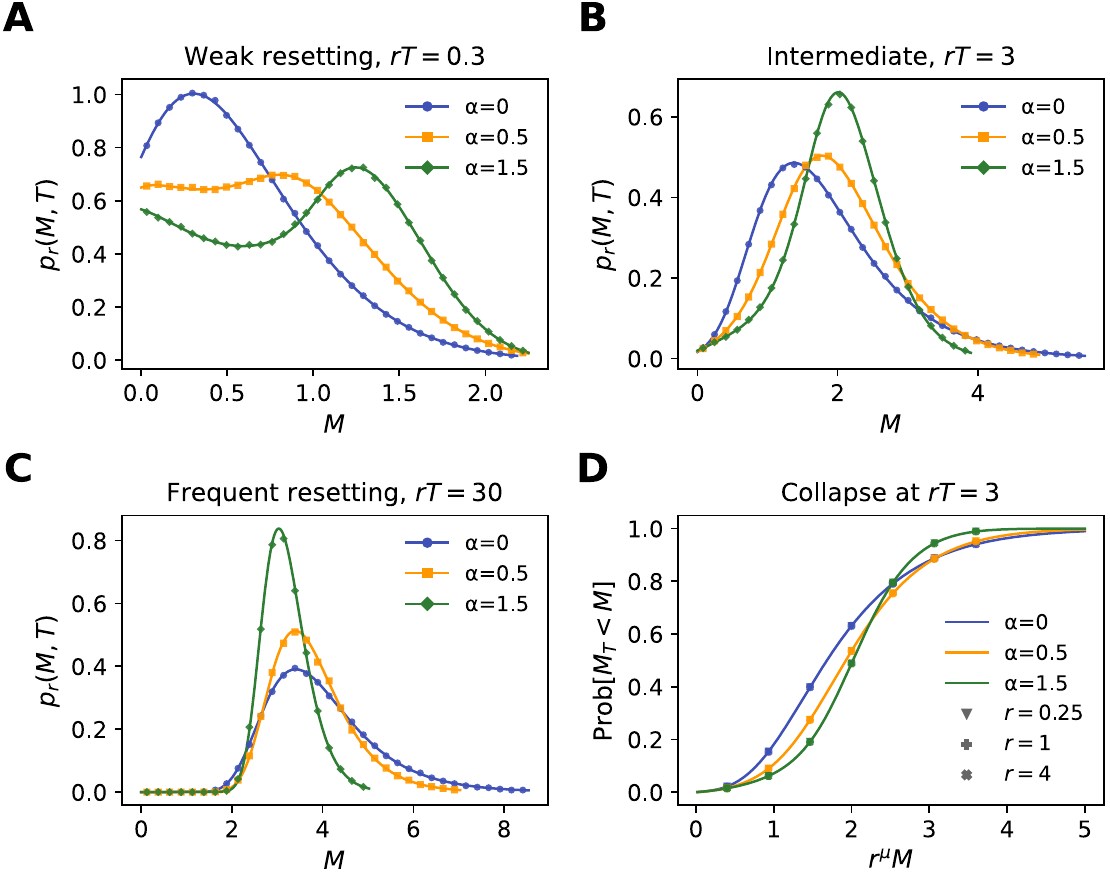}
\caption{Distribution of the maximum $M_T$ of the sluggish random walk with Poissonian resetting. \textbf{(A)}--\textbf{(C)} Probability density $p_r(M,T)$ at $r=1$ and weak ($rT=0.3$), intermediate ($rT=3$) and frequent ($rT=30$) resetting, for sluggishness parameters $\alpha=\{0, 0.5, 1.5\}$ (see legend). \textbf{(D)} Collapse of the distribution function $\mathrm{Prob}[M_T<M]$ at $rT=3$ as a function of $r^{\mu}M$, for rates $r=\{0.25, 1, 4\}$ (see legend). Symbols are obtained from numerical simulations started at the origin, with $N=10^6$ simulated trajectories per curve, generated in 10 independent blocks, with error bars given by the standard error across blocks, and lines are given by the numerical inversion of Eq.~\eqref{eq:Qr} (at $r=1$ in \textbf{(D)}).}
\label{fig:maxdist}
\end{figure}

Figure~\ref{fig:maxdist} illustrates these features. At weak resetting the sluggish densities are finite at the origin, as Eq.~\eqref{eq:p0} requires. For the larger sluggishness they are nonmonotonic, with a shallow dip that separates the origin from the main maximum, and the dip is absent for Brownian motion. As $rT$ grows the density moves to larger $M$ and the region near the origin empties, because the maximum becomes the largest of many interval maxima. The width first grows and, for $\alpha>0$, eventually decreases, whereas for Brownian motion it approaches a nonzero constant (Sec.~\ref{subsec:moments}). The ordering of the typical maxima with $\alpha$ reverses along the way. At intermediate resetting the more sluggish walks reach larger maxima than Brownian motion, and at frequent resetting the most sluggish walk has the smallest typical maximum. This reversal follows from the diffusivity profile. In the present units $D(x)=|x|^{-\alpha}$ exceeds the Brownian value $1$ for $|x|<1$ and falls below it for $|x|>1$, so sluggish walks are faster than Brownian ones at short distances. Changing the length unit moves the crossing point, and with it the value of $rT$ at which the ordering reverses, so the reversal point has no meaning independent of the units. At large $rT$ the typical maximum, Eq.~\eqref{eq:Mtyp}, grows as the power $2\mu<1$ of $\ln(rT)$ and eventually falls below the Brownian value, which holds in any units because only prefactors depend on the convention. The distribution function collapses when plotted against $r^{\mu}M$ for different rates $r$, as Eq.~\eqref{eq:Qr_scaling} implies (Fig.~\ref{fig:maxdist}D), while the curves for different $\alpha$ remain distinct, because $\alpha$ enters both through the relation between $Y$ and $M$ and through $G_\mu$.

\subsection{Long-time limit: Gumbel statistics}
\label{subsec:max_gumbel}

At long times the trajectory contains many resets, and the maximum is the largest of many independent interval maxima, which is the setting of classical extreme-value statistics. We therefore take the limit $rT\to\infty$ at fixed $r>0$, with the level $M=M(T)$ in the extreme-value scaling regime, and obtain from the exact transform~\eqref{eq:Qr} the Gumbel law of the maximum, its centering and its width in physical space.

\subsubsection{Dominant pole and the independent-intervals picture}
\label{subsubsec:pole}

At large $T$ the inverse Laplace transform of Eq.~\eqref{eq:Qr} is dominated by the pole $s_0(M)<0$ closest to the origin of the complex $s$-plane. It solves $s_0+r\,G_\mu(M,s_0+r)=0$. For large $M$ the function $G_\mu$ is small, and to leading order
\begin{equation}
s_0\simeq-\lambda(M),\qquad \lambda(M)=r\,G_\mu(M,r),\qquad \mathrm{Prob}[M_T<M]\simeq\exp\bigl[-\lambda(M)\,T\bigr].
\label{eq:pole}
\end{equation}
The pole is unique in the half-plane $\mathrm{Re}\,s>-r$ (Appendix~\ref{app:pole}). Equation~\eqref{eq:pole} has a direct interpretation. The trajectory consists of about $rT$ independent intervals, each of exponential duration with mean $1/r$. Each interval reaches the level $M$ with probability $G_\mu(M,r)$ (Sec.~\ref{subsec:survival_level}), so $\lambda(M)$ is the rate at which intervals that reach $M$ occur. The global maximum is the largest of the interval maxima, which is the setting of the classical extreme-value theory of independent variables~\cite{gumbel2004statistics,majumdar2020extreme}.

\subsubsection{Gumbel law in the Bessel variable}
\label{subsubsec:gumbel_y}

The large-argument form of $g_\mu$ gives, with $\zeta=\sqrt r\,Y$,
\begin{equation}
G_\mu(M,r)\simeq A_\mu\,\zeta^{1/2-\mu}e^{-\zeta}\Bigl[1+\frac{4\mu^2-1}{8\zeta}+\dots\Bigr],\qquad A_\mu=\frac{2^{\mu}\sqrt{\pi/2}}{\Gamma(1-\mu)} .
\label{eq:G_asym}
\end{equation}
Here $A_{1/2}=1$ and the correction in the bracket vanishes for $\alpha=0$. Inserting the leading term into Eq.~\eqref{eq:pole} gives $\mathrm{Prob}[\zeta_{\max}<\zeta]\simeq\exp\bigl[-A_\mu rT\,\zeta^{1/2-\mu}e^{-\zeta}\bigr]$, where $\zeta_{\max}=\sqrt r\,y(M_T)$. Writing $\zeta=a_T+\xi$ and solving $A_\mu rT\,\zeta^{1/2-\mu}e^{-\zeta}=e^{-\xi}$ by iteration gives $\mathrm{Prob}[\zeta_{\max}<a_T+\xi]\to\exp(-e^{-\xi})$, with
\begin{equation}
a_T=\ln\bigl(A_\mu rT\bigr)+\Bigl(\tfrac12-\mu\Bigr)\ln\ln\bigl(A_\mu rT\bigr)+o(1).
\label{eq:aT}
\end{equation}
The centered variable $\zeta_{\max}-a_T$ thus converges to the standard Gumbel law with centering $a_T$ and unit width~\cite{gumbel2004statistics,majumdar2020extreme}. Its mean is $\langle\zeta_{\max}\rangle=a_T+\gamma_E+o(1)$ for $rT\to\infty$, where $\gamma_E$ is the Euler constant. Equation~\eqref{eq:aT} contains the effect of sluggishness. For $\alpha=0$ the double-logarithmic term vanishes and $a_T=\ln(rT)$, which is the known Brownian result~\cite{singh2021extremal}. For $\alpha>0$ the term $(\tfrac12-\mu)\ln\ln(A_\mu rT)$ has no Brownian counterpart. It comes from the power-law prefactor $\zeta^{1/2-\mu}$ of $G_\mu$. Linearizing $\zeta^{1/2-\mu}$ around $a_T$ leaves corrections of relative order $1/a_T$, which decay only logarithmically in $T$, so the form $\exp[-rT\,G_\mu(M,r)]$ of Eq.~\eqref{eq:pole} is the more accurate approximation at finite $rT$. The data for the three values of $\alpha$ collapse onto the Gumbel law (Fig.~\ref{fig:gumbel}C).

\subsubsection{Typical maximum and width in physical space}
\label{subsubsec:gumbel_x}

Mapping back with $M=\mathcal X(Y)$ gives the typical maximum
\begin{equation}
M_{\rm typ}\simeq\left[\frac{(\alpha+2)\,a_T}{2\sqrt r}\right]^{2/(\alpha+2)}\simeq\left[\frac{(\alpha+2)\ln(rT)}{2\sqrt r}\right]^{2/(\alpha+2)} .
\label{eq:Mtyp}
\end{equation}
The second form keeps only the leading term of $a_T$. The double-logarithmic term $(\tfrac12-\mu)\ln\ln(A_\mu rT)$ decays only slowly relative to $\ln(rT)$, so the first form is the more accurate one at finite $rT$. The fluctuations of $\zeta_{\max}$ are of order one, and a fluctuation $\delta\zeta$ of the rescaled level produces a change $\delta M=2\mu M\,\delta\zeta/\zeta$ of the maximum. Since $M\propto\zeta^{2\mu}$, the Gumbel scale of $M_T$ is $b_T=2\mu M_{\rm typ}/a_T$, and the standard deviation of $M_T$ is
\begin{equation}
\sigma_T\simeq\frac{\pi}{\sqrt6}\,b_T=\frac{\pi}{\sqrt6}\,\frac{2\mu\,M_{\rm typ}}{a_T}\ \propto\ r^{-\mu}\bigl[\ln(rT)\bigr]^{2\mu-1}=r^{-\mu}\bigl[\ln(rT)\bigr]^{-\alpha/(\alpha+2)} .
\label{eq:bT}
\end{equation}
Because $b_T\ll M_{\rm typ}$, the map from $\zeta$ to $M$ is linear over the width of the distribution, so $(M_T-M_{\rm typ})/b_T$ is also Gumbel distributed. For $\alpha=0$ the standard deviation tends to the constant $\pi/\sqrt{6r}$ as $rT\to\infty$, independent of $T$. For every $\alpha>0$ it shrinks slowly with $T$. The origin of this narrowing is the sluggishness itself. A unit change of $Y$ corresponds to $dM=M^{-\alpha/2}\,dY$, which decreases with $M$, so that fluctuations of order one in the Bessel variable translate into ever smaller fluctuations of the position as the maximum grows.

\subsection{Moments and the crossover between free and reset regimes}
\label{subsec:moments}

The exact law also gives the moments of $M_T$. In particular, the mean follows from $\langle M_T\rangle=\int_0^\infty dM\,\mathrm{Prob}[M_T>M]$ and Eq.~\eqref{eq:Qr}, which give
\begin{equation}
\int_0^\infty dT\,e^{-sT}\langle M_T\rangle
=\frac{s+r}{s}\int_0^\infty dM\,\frac{G_\mu(M,s+r)}{s+rG_\mu(M,s+r)},
\label{eq:mean_max}
\end{equation}
and the second moment has an analogous expression. For $rT\ll1$ the mean grows as $\langle M_T\rangle\simeq C_\mu T^{\mu}$ with the amplitude
\begin{equation}
C_\mu=\frac{(2\mu)^{1-2\mu}}{\Gamma(1+\mu)}\int_0^\infty z^{2\mu-1}\,\frac{c_\mu}{g_\mu(z)}\,dz ,
\label{eq:Cmu}
\end{equation}
which equals $2/\sqrt\pi$ for $\alpha=0$. For $rT\gg1$ it grows as a power of $\ln(rT)$, $\langle M_T\rangle\simeq\mathcal X\bigl((a_T+\gamma_E)/\sqrt r\bigr)$, with the map $\mathcal X$ of Eq.~\eqref{eq:xinv} and $a_T$ from Eq.~\eqref{eq:aT}. This growth is logarithmic only for $\alpha=0$.

\begin{figure}
\centering
\includegraphics[width=\linewidth]{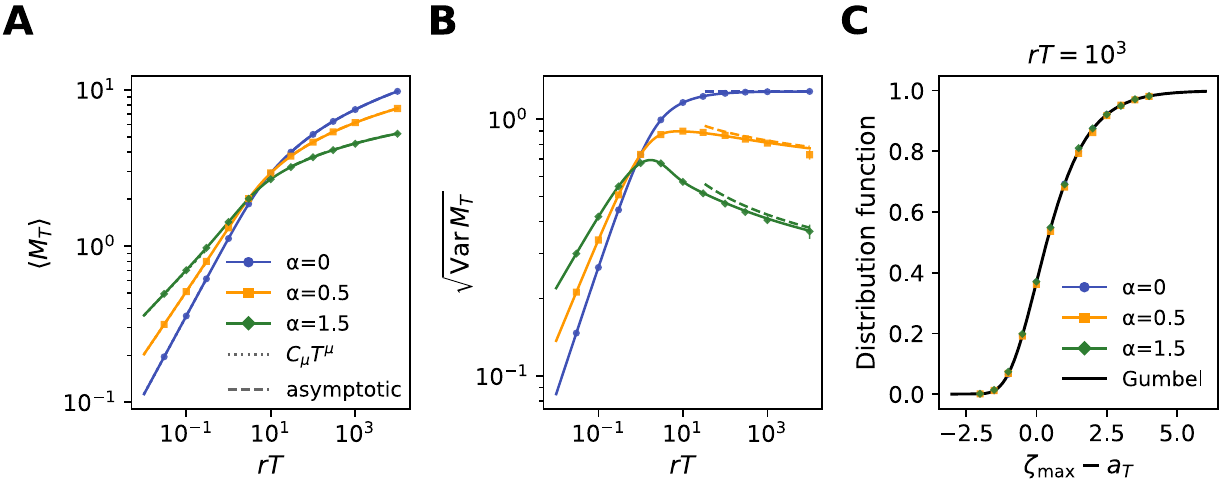}
\caption{Moments and Gumbel limit of the maximum $M_T$ at $r=1$. \textbf{(A)} Mean and \textbf{(B)} standard deviation of $M_T$ as functions of $rT$, and \textbf{(C)} distribution function of the centered variable $\zeta_{\max}-a_T$ at $rT=10^3$, for $\alpha=0$, $0.5$ and $1.5$ (see legend). Symbols are obtained from $N=10^6$ simulated trajectories per point, generated in 10 independent blocks, with error bars given by the standard error across blocks. Solid lines are given by the numerical inversion of Eq.~\eqref{eq:mean_max} and of its second-moment analogue in \textbf{(A)} and \textbf{(B)}, and by the standard Gumbel law in \textbf{(C)}. Dotted lines show the reset-free growth $C_\mu T^{\mu}$, Eq.~\eqref{eq:Cmu}, and dashed lines the large-$rT$ law $\mathcal X\bigl((a_T+\gamma_E)/\sqrt r\bigr)$ in \textbf{(A)} and the Gumbel width, Eq.~\eqref{eq:bT}, in \textbf{(B)}.}
\label{fig:gumbel}
\end{figure}

The mean and the standard deviation cross over between the two regimes (Figs.~\ref{fig:gumbel}A and~\ref{fig:gumbel}B). The standard deviation grows in the reset-free regime for every $\alpha$. For $\alpha=0$ it then rises monotonically to a plateau. For $\alpha>0$ it peaks at intermediate $rT$ and decreases at large $rT$, in agreement with Eq.~\eqref{eq:bT}. The Gumbel prediction approaches the exact curve only slowly. The mean curves for different $\alpha$ cross at intermediate $rT$, consistent with the reversal of the ordering of typical maxima discussed above.

\subsection{Far tail and a dynamical transition}
\label{subsec:max_ldp}

Equation~\eqref{eq:pole} describes the bulk of the distribution of the maximum. We next consider its far right tail. For levels far above the Gumbel scale, $\ln(rT)\ll\zeta\ll rT$, Eqs.~\eqref{eq:pole} and~\eqref{eq:G_asym} give the tail
\begin{equation}
\mathrm{Prob}[M_T>M]\simeq rT\,G_\mu(M,r)\simeq A_\mu\,rT\,\zeta^{1/2-\mu}e^{-\sqrt r\,Y} .
\label{eq:tail}
\end{equation}
The first-order form requires $rT\,G_\mu(M,r)\ll1$, which holds when $\zeta-a_T\to+\infty$.
For levels with $Y$ of order $T$ a second mechanism competes with this one, namely a single interval that lasts a time of order $T$. This large-deviation regime, in which $Y\sim T$, differs from the Gumbel regime $Y\sim\ln T$ of the preceding subsections. We therefore fix the resetting rate $r>0$ and the speed $w>0$ of the level, set $Y=wT$ and let $T\to\infty$, and look for the large-deviation rate function $\Phi(w)$ defined by $\mathrm{Prob}[y(M_T)>wT]\asymp e^{-T\Phi(w)}$~\cite{touchette2009large,meylahn2015large}.

Let $N$ be the number of inter-reset intervals whose maximum exceeds $M$. The union bound gives $\mathrm{Prob}[M_T>M]\le\langle N\rangle$, with
\begin{equation}
\langle N\rangle=\int_0^T d\tau\,\bigl[1+r(T-\tau)\bigr]\,e^{-r\tau}F_0(M,\tau).
\label{eq:union}
\end{equation}
Here $F_0(M,\tau)$ is the free first-passage density to $M$, so $e^{-r\tau}F_0(M,\tau)$ is the density of reaching $M$ at time $\tau$ after the start of an interval that has not yet ended. The factor $1+r(T-\tau)$ is the mean number of intervals that start early enough to leave a time $\tau$ before the end of the window, namely the interval that starts at the origin plus the intervals that start at the reset times in $[0,T-\tau]$. Equation~\eqref{eq:union} is also the first-order term of the expansion of Eq.~\eqref{eq:Qr} in $G_\mu$ (Appendix~\ref{app:ld}).

In the variable $y$ the free walker has unit diffusion constant and a drift that decays as $1/y$. Its first-passage time $\sigma_Y$ to the level $Y$ satisfies, uniformly in $Y$ and $\tau$, the Gaussian bounds
\begin{equation}
\frac{c}{1+Y/\sqrt\tau}\,e^{-Y^2/4\tau}\;\le\;\mathrm{Prob}[\sigma_Y\le\tau]\;\le\;C_\varepsilon\,e^{-(1-\varepsilon)Y^2/4\tau},
\label{eq:fpt_bounds}
\end{equation}
for every $\varepsilon\in(0,1)$, with constants $c>0$ and $C_\varepsilon<\infty$ that do not depend on $Y$ and $\tau$ (Appendix~\ref{app:ld}). For $Y=wT$ and $\tau=vT$ the free first-passage probability therefore has the exponential rate $w^2/(4v)$, uniformly in $v\in(0,1]$. This uniformity is what controls the integral in Eq.~\eqref{eq:union} near its minimizer and near the endpoints $\tau\to0$ and $\tau=T$. Integrating Eq.~\eqref{eq:union} by parts and inserting the upper bound gives $\langle N\rangle\le C'_\varepsilon(1+rT)^2\max_{0<\tau\le T}\exp[-r\tau-(1-\varepsilon)Y^2/4\tau]$, and $\varepsilon\to0$ is taken at the end. The maximum is attained at the value of $\tau$ that minimizes the exponent. This is $\tau^*=Y/(2\sqrt r)$ when $\tau^*<T$, and $\tau=T$ otherwise. A matching lower bound follows from the event that the initial interval alone reaches $M$ before the time $\tau_1=\min(\tau^*,T)$ without having been reset. For $w>2\sqrt r$ one has $\tau_1=T$, and this is the event of no reset, which has probability $e^{-rT}$ and reaches $Y$ with a probability of exponential rate $Y^2/4T$ (Appendix~\ref{app:ld}). Evaluating the exponent at the two minima gives
\begin{equation}
\Phi(w)=\begin{cases}\sqrt r\,w, & w<2\sqrt r,\\[6pt] \dfrac{w^2}{4}+r, & w>2\sqrt r .\end{cases}
\label{eq:ldp}
\end{equation}
The prefactors in Eqs.~\eqref{eq:tail} and~\eqref{eq:union} grow only algebraically in $T$. At $Y=wT$ they are subleading compared with the exponential factor, so the first branch of Eq.~\eqref{eq:ldp} agrees with Eq.~\eqref{eq:tail}. The rate function does not depend on $\alpha$ in the variable $Y$, because only the prefactors do. For $\alpha=0$ Eq.~\eqref{eq:ldp} is the rate function of the maximum of Brownian motion with resetting, whose second-order singularity was found in Ref.~\cite{hartmann2024distribution}. What is new here is that the same function, with the same transition, governs the maximum for every $\alpha\ge0$ in the variable $Y$.

In physical space write $M=m\,T^{2/(\alpha+2)}$, where $m$ is the scaled maximum, so that $Y=wT$ with $w=2m^{(\alpha+2)/2}/(\alpha+2)$. Then $\mathrm{Prob}[M_T>m\,T^{2/(\alpha+2)}]\asymp e^{-T\mathcal I_\alpha(m)}$ with the rate function
\begin{equation}
\mathcal I_\alpha(m)=\begin{cases}\dfrac{2\sqrt r}{\alpha+2}\,m^{(\alpha+2)/2}, & m<m_c,\\[8pt] r+\dfrac{m^{\alpha+2}}{(\alpha+2)^2}, & m>m_c,\end{cases}\qquad m_c=\bigl[(\alpha+2)\sqrt r\bigr]^{2/(\alpha+2)} .
\label{eq:ldp_phys}
\end{equation}
For $\alpha=0$ this is $\sqrt r\,m$ for $m<2\sqrt r$ and $r+m^2/4$ beyond. The transition therefore sits at the level $M_c=m_c\,T^{2/(\alpha+2)}$, and sluggishness enters the far tail only through the exponent $2/(\alpha+2)$ of the scaling of $M$ with $T$ and through the shape of $\mathcal I_\alpha$ in $m$.

At $w_c=2\sqrt r$ both branches equal $2r$ and both slopes equal $\sqrt r$, while the second derivative jumps from $0$ to $1/2$. The rate function and its first derivative are therefore continuous and its second derivative is discontinuous. This is a dynamical transition of the rate function, of second order in the Ehrenfest sense, and not a phase transition of the process. It reflects a change of the dominant mechanism. For $0<w<w_c$ the dominant event is a single anomalously long interval of duration $\tau^*=Y/(2\sqrt r)=(w/w_c)\,T<T$, which climbs to the level $Y$ and ends by a reset before the window does. The reset typically occurs within a time of order $1/r$ after the level is reached, which is subleading on the scale $T$. For $w>w_c$ the optimal duration $\tau^*$ would exceed $T$, and the dominant event is an interval that occupies essentially the whole window. The trajectories without a reset provide the lower bound, and histories with a reset close to the start of the window have the same exponential rate. A related transition occurs in the relaxation of reset processes to their stationary state~\cite{majumdar2015dynamical}.

\begin{figure}
\centering
\includegraphics[width=\linewidth]{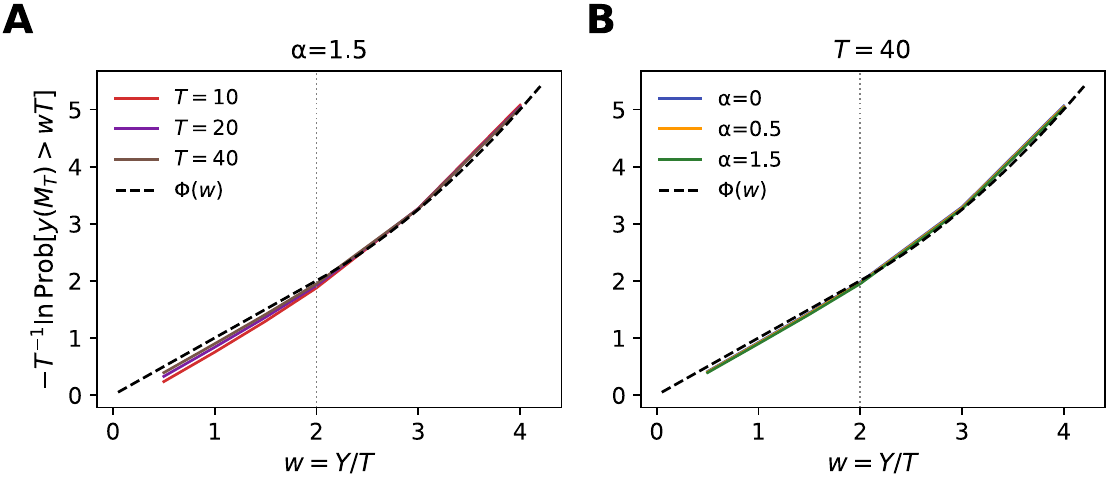}
\caption{Large deviations of the maximum $M_T$ at $r=1$. \textbf{(A)} Rate $-T^{-1}\ln\mathrm{Prob}[y(M_T)>wT]$ for $\alpha=1.5$ at $T=\{10, 20, 40\}$ (see legend), from the numerical inversion of Eq.~\eqref{eq:Qr} in extended precision, approaching the rate function $\Phi(w)$, Eq.~\eqref{eq:ldp} (dashed line). The dotted vertical line marks $w_c=2\sqrt r$. \textbf{(B)} The same rate at $T=40$ for $\alpha=\{0, 0.5, 1.5\}$ (see legend).}
\label{fig:ld}
\end{figure}

The finite-time rate curves of Fig.~\ref{fig:ld}A approach $\Phi(w)$ as $T$ grows, from below for $w<w_c$. The rate increases linearly with $w$ up to the transition and bends upwards quadratically beyond it.

\section{Time of the maximum}
\label{sec:tmax}

\subsection{Path decomposition and the joint law of the maximum and its time}
\label{subsec:tmax_decomp}

We now turn to the time at which the maximum occurs. Its distribution follows from the joint distribution of the maximum $M$ and its time $t_M$, which we obtain by splitting each trajectory at $t_M$. Before $t_M$ the walker stays below $M$ and reaches $M$ at time $t_M$. After $t_M$ it starts just below $M$ and stays below $M$ for the remaining time $T-t_M$. Both pieces evolve under resetting, and since the process is Markov, the two pieces are independent given $M$ and $t_M$. The second piece starts at the absorbing boundary $x=M$, where the first piece ends. The absorbing-boundary limit is taken by starting the second piece at $M-\varepsilon$ and letting $\varepsilon\to0$ after normalization~\cite{majumdar2004exact,majumdar2005airy,randon2007distribution,majumdar2008time,del2025generalized}. In the Bessel variable the diffusion constant is one and the argument used for Brownian motion applies near the level $Y$. The argmax $t_M$ is almost surely unique, because a time-changed Brownian path attains its maximum at a single time with probability one. The argument gives the joint density of $(Y,t_M)$ as the product of the first-passage density to $Y$ and the slope $-\partial_{y_0}Q$ of the survival probability at $Y$. Transforming back to $M$ with $dY=M^{\alpha/2}dM$ gives
\begin{equation}
P(M,t_M|T)=F_r(M,t_M)\;q_r(M,T-t_M).
\label{eq:joint}
\end{equation}
Here $F_r(M,t)$ is the first-passage density to $M$ from the origin under resetting, and $q_r(M,\tau)=-\partial_{x_0}Q_r(x_0,\tau\,|\,M)\big|_{x_0=M}$ is the slope at the absorbing level of the survival probability $Q_r(x_0,\tau\,|\,M)$ of the reset process started below $M$. A derivation from the strong Markov property is given in Appendix~\ref{app:decomp}.

The two factors have simple Laplace transforms. Let $s_1$ and $s_2$ be the Laplace variables conjugate to $t_M$ and $T-t_M$. Writing $u_i=s_i+r$ for the shifted variables and $G_i=G_\mu(M,u_i)$, Eq.~\eqref{eq:Qr} gives
\begin{equation}
\tilde F_r(M,s_1)=1-s_1\tilde Q_r(M,s_1)=\frac{u_1\,G_1}{s_1+r\,G_1} .
\label{eq:Fr}
\end{equation}
For the second factor we need the slope of the free survival probability at the absorbing level. Differentiating Eq.~\eqref{eq:Q0_general} at $x_0=M$ and using $dY/dM=M^{\alpha/2}$ gives $\tilde q_0(M,s)=M^{\alpha/2}\eta(\sqrt s\,Y)/\sqrt s$. Differentiating Eq.~\eqref{eq:renewal_b_LT} at $x_0=M$, where only the first factor depends on $x_0$, then gives
\begin{equation}
\tilde q_r(M,s_2)=\tilde q_0(M,u_2)\bigl[1+r\tilde Q_r(M,s_2)\bigr]=\frac{u_2}{s_2+r\,G_2}\;\frac{M^{\alpha/2}}{\sqrt{u_2}}\,\eta\bigl(\sqrt{u_2}\,Y\bigr) .
\label{eq:qr}
\end{equation}
We define the double Laplace transform of the joint density in $t_1=t_M$ and $t_2=T-t_M$,
\begin{equation}
\hat P(M,s_1,s_2)=\int_0^\infty dt_1\int_0^\infty dt_2\,e^{-s_1t_1-s_2t_2}\,P(M,t_1|t_1+t_2)=\tilde F_r(M,s_1)\,\tilde q_r(M,s_2).
\label{eq:Phat}
\end{equation}
The segment after the maximum restarts at $M$ but resets to the origin.

\subsection{Distribution of the time of the maximum}
\label{subsec:tmax_marginal}

Integrating Eq.~\eqref{eq:Phat} over $M$ and changing variable to $Y$ gives the main result of this section,
\begin{equation}
\hat p(s_1,s_2)=\int_0^\infty dY\;\frac{u_1\,G_\mu(M,u_1)}{s_1+r\,G_\mu(M,u_1)}\;\frac{\sqrt{u_2}\;\eta\bigl(\sqrt{u_2}\,Y\bigr)}{s_2+r\,G_\mu(M,u_2)} ,
\label{eq:ptilde}
\end{equation}
where $G_\mu(M,u)=c_\mu/g_\mu(\sqrt u\,Y)$ and $M=\mathcal X(Y)$. The two factors of the integrand are the Laplace transforms of the first-passage density $F_r(Y,t)$ to the level $Y$ and of the kernel $K_r(Y,t)$ of the segment after the maximum. In this form the sluggishness enters only through the order $\mu$ of the Bessel functions, and for $\alpha=0$ Eq.~\eqref{eq:ptilde} reduces to the Brownian double Laplace transform of Ref.~\cite{singh2021extremal}. We note three properties of Eq.~\eqref{eq:ptilde}.

\textit{Normalization.} For $s_1=s_2=s$ the substitution $\chi=g_\mu(\sqrt{u}\,Y)$ gives
\begin{equation}
\hat p(s,s)=\int_{c_\mu}^{\infty}\frac{u\,c_\mu\,d\chi}{(s\,\chi+r\,c_\mu)^2}=\frac1s ,
\label{eq:norm}
\end{equation}
which is the Laplace transform of $\int_0^Tp(t_M|T)\,dt_M=1$.

\textit{Reset-free limit.} For $r=0$ and $\alpha=0$, Eq.~\eqref{eq:ptilde} gives $\hat p(s_1,s_2)=1/\sqrt{s_1s_2}$, the double Laplace transform of L\'evy's arcsine law. For general $\alpha$ at $r=0$, writing $p(t_M|T)=T^{-1}f_M(t_M/T)$, where $f_M$ is the scaled density, gives the Stieltjes representation
\begin{equation}
\int_0^1\frac{f_M(\xi)}{z-\xi}\,d\xi
=\frac{c_\mu}{z}\int_0^\infty dx\,
\frac{g_\mu'(x)}{g_\mu(x)g_\mu(\sqrt{1-1/z}\,x)},\qquad z>1,
\label{eq:stieltjes}
\end{equation}
which is the known law of the free sluggish walk in the present notation~\cite{del2025generalized}. It implies the edge behavior $f_M(\xi)\sim\xi^{\mu-1}$ for $\xi\to0$ and $f_M(\xi)\sim(1-\xi)^{-1/2}$ for $\xi\to1$, in agreement with Ref.~\cite{del2025generalized}, and it reduces to the arcsine law at $\alpha=0$.

\textit{Finite-time scaling.} Equation~\eqref{eq:ptilde} depends on $s_1,s_2$ only through $s_i/r$, apart from an overall factor $1/r$, which gives the scaling form $p(t_M|T)=T^{-1}\mathcal F_\alpha(t_M/T,rT)$. We compute $\mathcal F_\alpha$ in the time domain. Equation~\eqref{eq:joint} gives $p(t_M|T)=\int dY\,F_r(Y,t_M)\,K_r(Y,T-t_M)$, where $F_r(Y,t)$ and $K_r(Y,t)$ are the inverse Laplace transforms of $\tilde F=uG/(s+rG)$ and $K=\sqrt u\,\eta/(s+rG)$, with $u=s+r$. This requires only one-dimensional inversions followed by a quadrature over $Y$ (Appendix~\ref{app:numerics}). The resulting density integrates to unity.

\begin{figure}
\centering
\includegraphics[width=\linewidth]{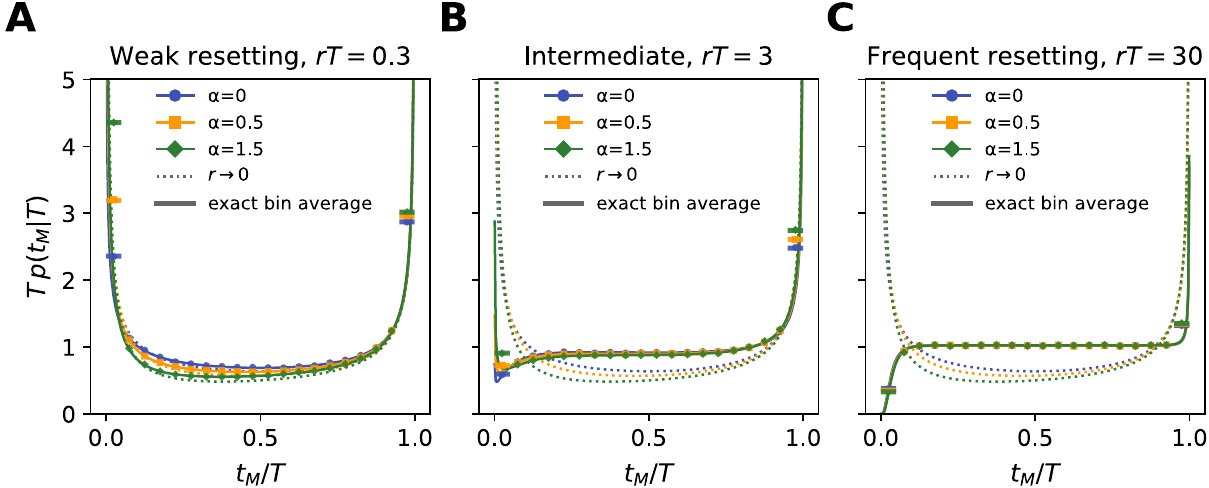}
\caption{Distribution of the time of the maximum, $T\,p(t_M|T)$, as a function of $t_M/T$ at weak ($rT=0.3$, \textbf{(A)}), intermediate ($rT=3$, \textbf{(B)}) and frequent ($rT=30$, \textbf{(C)}) resetting, for sluggishness parameters $\alpha=\{0, 0.5, 1.5\}$ (see legend). Symbols are averages over 20 equal bins of $N=10^6$ simulated trajectories, generated in 10 independent blocks, with error bars given by the standard error across blocks. In the two outermost bins, where the density diverges, the horizontal bars show the exact bin averages obtained by integrating the density over the bin. Solid lines are given by Eq.~\eqref{eq:ptilde} in the time domain. Dotted lines show the reset-free law ($r\to0$) of the same color, which is the arcsine law for $\alpha=0$.}
\label{fig:tMdist}
\end{figure}

Figure~\ref{fig:tMdist} shows the evolution of the density with $rT$. At weak resetting the law remains close to the free U-shaped distribution (dotted), with singular behavior at both endpoints. For $\alpha=0$ the free law is the symmetric arcsine law, whereas for $\alpha>0$ it is asymmetric and carries more weight at early times. As $rT$ grows the weight near $t_M=0$ is strongly suppressed and the bulk becomes flat, as shown analytically below, while the divergence at $t_M=T$ persists and narrows. At any finite $rT$ the density diverges as $t_M^{\mu-1}$ for $t_M\to0$ with an amplitude proportional to $e^{-rT}$, because trajectories without a reset, which have probability $e^{-rT}$, retain the free-law edge. This divergence is still visible at intermediate resetting, where the bulk is also tilted toward late times. At frequent resetting it is invisible on the scale of the figure, the three curves are nearly indistinguishable in the bulk, and they differ only close to the endpoints.

\subsection{Long-time behavior: uniform bulk and boundary layers}
\label{subsec:tmax_longtime}

At long times the integral in Eq.~\eqref{eq:ptilde} is dominated by large $Y$, where $G\equiv G_\mu(M,r)$ is small. For $s\ll r$ and large $Y$, Eq.~\eqref{eq:G_asym} gives
\begin{equation}
G_\mu(M,r+s)\simeq G_\mu(M,r)\,e^{-s\,\tau_Y},\qquad \tau_Y=\frac{Y\,\eta(\sqrt r\,Y)}{2\sqrt r},
\label{eq:delay}
\end{equation}
since $\partial_s\ln G_\mu(M,r+s)\big|_{s=0}=-Y\eta/(2\sqrt r)$. For $\alpha=0$ one has $\eta=1$ and $\tau_Y=Y/(2\sqrt r)$. The factor $e^{-s\tau_Y}$ is the Laplace transform of a delay $\tau_Y$. To leading order it is the time that the optimal path of Sec.~\ref{subsec:max_ldp} needs to climb to the level $Y$ within one interval. At the typical maximum, $\sqrt r\,Y\simeq a_T$, this delay is $\tau_Y\simeq a_T/(2r)\simeq\ln(rT)/(2r)$.

When $t_M$ and $T-t_M$ are both much larger than $\tau_Y$, that is $s_i\tau_Y\ll1$, one may set $u_i\simeq r$ and drop the delay. Since $d\lambda/\lambda=-\sqrt r\,\eta\,dY$ for $\lambda=rG_\mu(M,r)$, the substitution to $\lambda$ turns Eq.~\eqref{eq:ptilde} into
\begin{equation}
\hat p(s_1,s_2)\simeq\int_0^{r}\frac{d\lambda}{(s_1+\lambda)(s_2+\lambda)}\simeq\frac{\ln(s_1/s_2)}{s_1-s_2} .
\label{eq:bulk}
\end{equation}
The right-hand side is the double Laplace transform of $1/(t_1+t_2)=1/T$. Hence, at long times, $t_M$ is uniformly distributed in the bulk of $[0,T]$ for every $\alpha$. For $\alpha=0$ this bulk law was obtained in Ref.~\cite{singh2021extremal}, and the boundary layers below are analyzed here. The bulk law agrees with the independent-intervals picture. The global maximum is reached in one of many statistically identical intervals, and that interval is equally likely to start anywhere in the window~\cite{singh2021extremal}.

The approximations $u_i\simeq r$ and the neglect of the delay fail near the two ends of the window, and the two ends are not alike. At the left end, a high maximum cannot be reached at once. Within the interval that produces it the walker needs the climbing time $\tau_Y$, of order $\ln(rT)/(2r)$, so $t_M$ is depleted near the start of the window out to about this time. At the right end the walker only has to stay below the maximum for the short remaining time $T-t_M$. This costs only a power $(T-t_M)^{-1/2}$, as for free diffusion. Once the remaining time exceeds the reset time $1/r$, a reset to the origin makes the constraint cost nothing more and the density is flat.

\textit{Right layer.} For $T-t_M$ of order $1/r$ the first factor of Eq.~\eqref{eq:ptilde} is the first-passage density to a high level at a late time, $F_r(Y,T)\simeq \lambda\,e^{-\lambda T}$ with the rate $\lambda=rG_\mu(M,r)$. At the levels that matter $G$ is small, and the transform of the second factor reduces to $\sqrt{s+r}/s$, whose inverse is $e^{-rt}/\sqrt{\pi t}+\sqrt r\,\mathrm{erf}\sqrt{rt}$, where $\mathrm{erf}$ is the error function. Since $\int dY\,\lambda\,e^{-\lambda T}=1/(\sqrt r\,T)$ to leading order (Appendix~\ref{app:layers}),
\begin{equation}
T\,p(t_M|T)\;\to\;\kappa(\rho)=\mathrm{erf}\sqrt \rho+\frac{e^{-\rho}}{\sqrt{\pi \rho}},\qquad \rho=r(T-t_M).
\label{eq:kappa}
\end{equation}
The layer has width of order $1/r$ and diverges as $(T-t_M)^{-1/2}$ at the end of the window. It carries the excess weight $\int_0^\infty(\kappa-1)\,d\rho/r=1/(2r)$. For $\alpha>0$ there are corrections of relative order $1/\ln(rT)$, which come from $\eta\neq1$.

\textit{Left layer.} For $t_M\ll T$ the first-passage density to $Y$ is, to first order in $G$, $F_r(Y,t)=h_Y(t)+r\int_0^th_Y$ with $h_Y(t)=e^{-rt}F_0(M,t)$, and the second factor is $\sqrt r\,\eta\,e^{-\lambda T}$. Let $\pi_Y$ be the density with Laplace transform $G_\mu(M,r+s)/G_\mu(M,r)$ and let $\Pi_Y$ be its distribution function. It is the law of the time that an interval needs to climb to the level $Y$, given that it reaches $Y$ before the next reset. The weight $T\sqrt r\,\eta\,\lambda\,e^{-\lambda T}dY$ is the distribution of the level $Y_{\max}$ of the global maximum, and therefore
\begin{equation}
T\,p(t_M|T)\;\simeq\;\Bigl\langle\,\Pi_Y(t_M)+\frac{\pi_Y(t_M)}{r}\Bigr\rangle_{Y_{\max}} .
\label{eq:leftlayer}
\end{equation}
The mean climbing time is $\tau_Y=-\partial_s\ln G_\mu(M,r+s)|_{s=0}=Y\eta/(2\sqrt r)$, which is the delay of Eq.~\eqref{eq:delay}, and its spread is of order $\sqrt{\ln(rT)}/r$. The left layer is therefore a depletion region centered near $\ln(rT)/(2r)$.

\begin{figure}
\centering
\includegraphics[width=\linewidth]{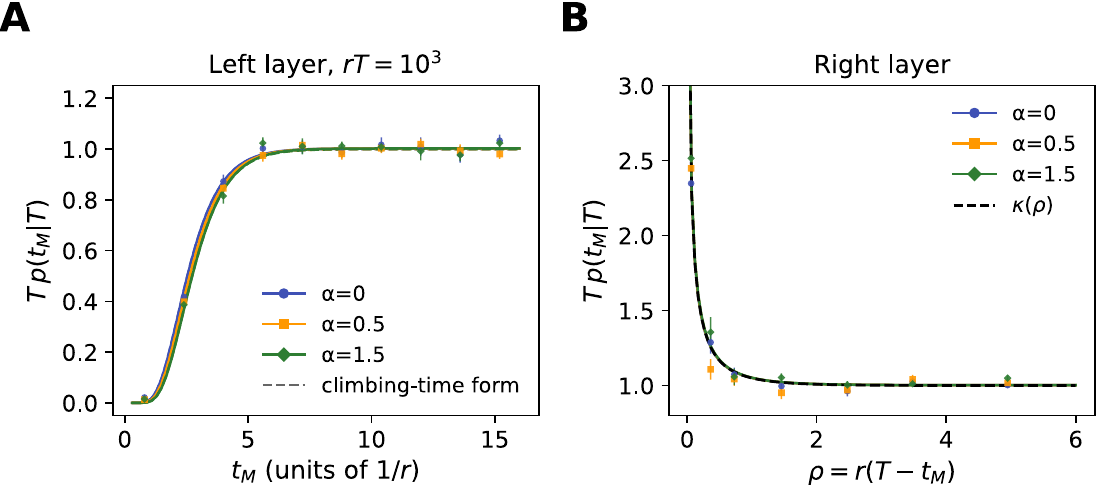}
\caption{Boundary layers of the time of the maximum at $rT=10^3$, for sluggishness parameters $\alpha=\{0, 0.5, 1.5\}$ (see legend). \textbf{(A)} Left layer: $T\,p(t_M|T)$ near $t_M=0$ as a function of $t_M$ in units of $1/r$. \textbf{(B)} Right layer: $T\,p(t_M|T)$ near $t_M=T$ as a function of $\rho=r(T-t_M)$. Symbols are binned averages of $N=10^6$ simulated trajectories, generated in 10 independent blocks, with error bars given by the standard error across blocks. Solid lines are given by the exact law, and dashed lines show the climbing-time form, Eq.~\eqref{eq:leftlayer}, in \textbf{(A)} and $\kappa(\rho)$, Eq.~\eqref{eq:kappa}, in \textbf{(B)}.}
\label{fig:layers}
\end{figure}

The two layers hold in different scaling regimes. Equation~\eqref{eq:kappa} holds at fixed $\rho=r(T-t_M)$ as $T\to\infty$. Equation~\eqref{eq:leftlayer} concerns the left boundary region $t_M=o(T)$, in which the climbing scale $\ln(rT)/r$ sets the extent of the depletion. Between the two the density approaches the flat bulk value of Eq.~\eqref{eq:bulk}. Both layers depend only weakly on $\alpha$ (Fig.~\ref{fig:layers}), so the sluggishness enters mainly through small shifts, and the right layer lies on the single curve $\kappa(\rho)$ for all three values. These layers can be compared with those of Brownian motion with resetting that starts from its stationary state~\cite{mori2022time}. There the edges of the density of $t_M$ are $G(rt_M)/T$ at the left and $[2G(\rho)-1]/T$ at the right, with $G(z)=\tfrac12\bigl[1+\mathrm{erf}\sqrt z+e^{-z}/\sqrt{\pi z}\bigr]$. Since $2G(\rho)-1=\kappa(\rho)$, the right layer of Eq.~\eqref{eq:kappa} is the same function, which is consistent with the fact that it depends only on the time of order $1/r$ before the end of the window and not on how the walk began. The left layers differ. A walker started in the stationary state is already at a typical position and has an excess of weight, of width $1/r$, near $t_M=0$. A walker started at the reset point must first climb to the level of the global maximum, which takes a time of order $\ln(rT)/(2r)$, so its left layer is a depletion whose extent grows with $T$. This is the origin of the logarithmic growth of the mean offset in Sec.~\ref{subsec:tmax_mean}. The stationary-start case does not include sluggishness.

\subsection{Mean time of the maximum}
\label{subsec:tmax_mean}

The two boundary layers remove weight from the left of the window and add weight at the right, so they displace the mean of $t_M$ from $T/2$. The density is uniform in the bulk and has a boundary layer at each end. The left layer removes the weight $W_L=\langle\tau_Y\rangle-1/r$ near $t_M=0$, because the maximum cannot be reached before the climbing time has elapsed. The right layer, which is confined to the reset scale $1/r$, adds the weight $W_R=1/(2r)$ near $t_M=T$ (Appendix~\ref{app:layers}). A weight $W$ added at $t_M\simeq T$ or removed at $t_M\simeq0$ shifts the mean by $W/2$. Normalization requires the net weight $W_L-W_R$ to be returned to the bulk, and since the bulk density is uniform this compensation is spread uniformly over $[0,T]$ and does not move the mean. The mean therefore shifts by $(W_L+W_R)/2$, so that $\langle t_M\rangle-T/2=\langle\tau_Y\rangle/2-1/(4r)$, and the leading behavior is $\ln(rT)/(4r)$. This weight counting is a heuristic, but it also reproduces the constants. By Eq.~\eqref{eq:delay} the mean climbing time is $\tau_Y=\zeta\eta(\zeta)/(2r)$ with $\zeta=\sqrt r\,Y$, and $\eta\simeq1-(\tfrac12-\mu)/\zeta$ at large $\zeta$. Averaging over the Gumbel law of $\zeta_{\max}$ gives $\langle\tau_Y\rangle=[a_T+\gamma_E-(\tfrac12-\mu)]/(2r)$, so that $\langle t_M\rangle-T/2=[a_T+\gamma_E+\mu-\tfrac32]/(4r)$, which is Eq.~\eqref{eq:tM_shift}. The exact derivation below does not rely on this argument.

Since $t_1=t_M$ and $t_2=T-t_M$, one has $\langle t_M\rangle-T/2=(\langle t_1\rangle-\langle t_2\rangle)/2$, and therefore
\begin{equation}
\int_0^\infty dT\,e^{-sT}\Bigl[\langle t_M\rangle_T-\frac T2\Bigr]=\frac12\bigl(\partial_{s_2}-\partial_{s_1}\bigr)\hat p(s_1,s_2)\Big|_{s_1=s_2=s} .
\label{eq:tM_mean}
\end{equation}
The right-hand side simplifies because $g_\mu$ obeys a second-order differential equation (Appendix~\ref{app:mean}), and one finds
\begin{equation}
\int_0^\infty dT\,e^{-sT}\Bigl[\langle t_M\rangle_T-\frac T2\Bigr]=\frac14\int_0^\infty dZ\;\frac{G(Z)\,\bigl[Z-d\,\eta(Z)\bigr]}{\bigl[s+r\,G(Z)\bigr]^2},
\label{eq:tM_exact}
\end{equation}
where $Z=\sqrt{s+r}\,Y$ is the rescaled level, $G(Z)=c_\mu/g_\mu(Z)$, $\eta(Z)=g_\mu'(Z)/g_\mu(Z)$ and $d=2(1-\mu)$. The sluggishness enters only through $G$, $\eta$ and $d$, and Eq.~\eqref{eq:tM_exact} is a single integral.

\begin{figure}
\centering
\includegraphics[width=\linewidth]{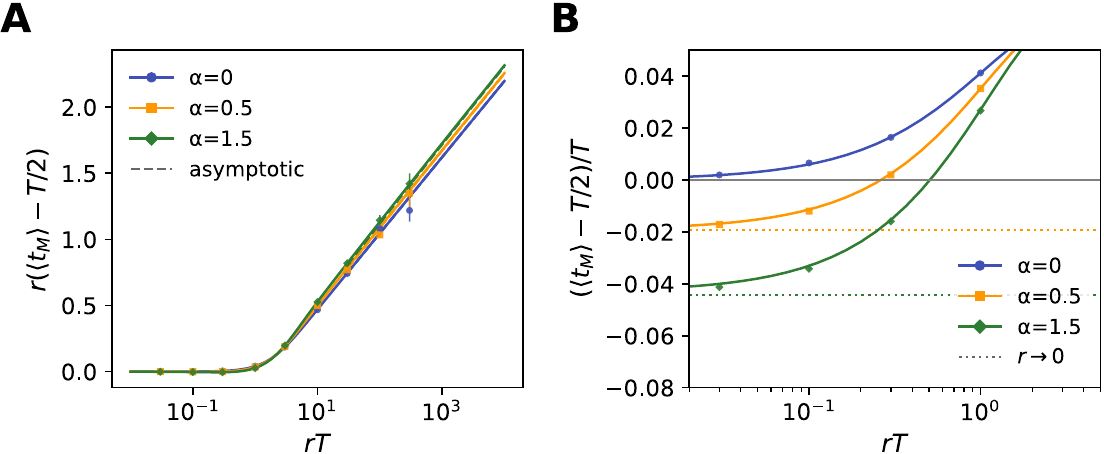}
\caption{Mean offset of the time of the maximum from $T/2$, for sluggishness parameters $\alpha=\{0, 0.5, 1.5\}$ (see legend). \textbf{(A)} $r(\langle t_M\rangle-T/2)$ as a function of $rT$. \textbf{(B)} $(\langle t_M\rangle-T/2)/T$ at small $rT$. Symbols are obtained from $N=10^6$ simulated trajectories, generated in 10 independent blocks, with error bars given by the standard error across blocks, and solid lines are given by Eq.~\eqref{eq:tM_exact} inverted numerically (the closed form~\eqref{eq:tM_alpha0} for $\alpha=0$). Dashed lines show the asymptotic law, Eq.~\eqref{eq:tM_shift}, in \textbf{(A)}, and dotted lines the reset-free values of the same color in \textbf{(B)} (zero for $\alpha=0$).}
\label{fig:offset}
\end{figure}

For $r\to0$ Eq.~\eqref{eq:tM_exact} reduces to the free sluggish-walk expression of Ref.~\cite{del2025generalized}. For $\alpha>0$ the free mean lies below $T/2$, at $0.4807\,T$ for $\alpha=0.5$ and at $0.4556\,T$ for $\alpha=1.5$. Resetting reverses the sign of the offset at $rT\simeq0.26$ and $0.52$, respectively, and the offset then grows logarithmically. For $\alpha=0$ the free mean is $T/2$ and the offset is positive for every $rT>0$. For $\alpha=0$ one has $G=e^{-Z}$, $\eta=1$ and $d=1$, and the integral is elementary. The transform equals $[\ln(1+r/s)-r/(s+r)]/(4rs)$, which inverts to
\begin{align}
    \langle t_M\rangle_T-\frac T2&=\frac{1}{4r}\Bigl[\mathrm{Ein}(rT)-1+e^{-rT}\Bigr],\notag\\[6pt] \mathrm{Ein}(x)&=\int_0^x\frac{1-e^{-t}}{t}\,dt=\ln x+\gamma_E+E_1(x),
\label{eq:tM_alpha0}
\end{align}
where $E_1(x)=\int_x^\infty e^{-t}\,dt/t$ is the exponential integral. This coincides with the Brownian result~\cite{singh2021extremal}. For $rT\ll1$ it gives $rT^2/16$, which vanishes at $r=0$. For $rT\gg1$ it gives $[\ln(rT)+\gamma_E-1]/(4r)$. For general $\alpha$ and $rT\gg1$, the small-$s$ expansion of the exact transform~\eqref{eq:tM_exact} (Appendix~\ref{app:mean}) gives, as an asymptotic inversion at fixed $r$ for $rT\to\infty$,
\begin{equation}
\langle t_M\rangle_T-\frac T2\simeq\frac{1}{4r}\Bigl[a_T+\gamma_E+\mu-\tfrac32\Bigr],
\label{eq:tM_shift}
\end{equation}
where $a_T$ is the centering of the Gumbel law, Eq.~\eqref{eq:aT}. Since $\langle\zeta_{\max}\rangle\simeq a_T+\gamma_E$, this equals $\bigl[\langle\zeta_{\max}\rangle-(\tfrac32-\mu)\bigr]/(4r)$. For $\alpha=0$ it reduces to Eq.~\eqref{eq:tM_alpha0}. The leading coefficient $1/4$ is the one predicted by the two boundary layers. It is independent of $\alpha$, so the offset grows as $\ln(rT)/(4r)$ for every $\alpha$. The terms $(\tfrac12-\mu)\ln\ln(A_\mu rT)$, $\ln A_\mu$ and $\mu-\tfrac32$ depend on $\alpha$. The offset is small at weak resetting, rises at intermediate resetting, and then grows as a straight line on the logarithmic axis (Fig.~\ref{fig:offset}A), approaching the asymptotic law~\eqref{eq:tM_shift}. The slope is nearly the same for all $\alpha$ and only a small vertical shift depends on $\alpha$, as expected from the $\alpha$-independent leading coefficient. Panel B shows the small-$rT$ regime. For the sluggish walks the offset starts negative because the free law favors early times, crosses zero at an $rT$ that is larger for larger $\alpha$, and then becomes positive. For Brownian motion it is positive from the outset.

\section{Last-passage time and a global-time symmetry}
\label{sec:lastpassage}

\subsection{Exact distribution of the last-passage time}
\label{subsec:tl_exact}

The last-passage time admits a simple decomposition because resets land on the origin. Let $\tau_{\rm last}$ be the time of the last reset before $T$, with $\tau_{\rm last}=0$ if no reset occurs. After $\tau_{\rm last}$ the walker is a free sluggish walker started at the origin and observed during the time $\tau=T-\tau_{\rm last}$. Its last zero in this window is $\tau_{\rm last}+\tau\,\xi$, where $\xi\in[0,1]$ is the scaled last-zero time of the free walker, which follows the free last-passage law~\cite{del2025generalized}
\begin{equation}
f_l(\xi)=\frac{\sin\pi\mu}{\pi}\,\frac{1}{\xi^{1-\mu}(1-\xi)^{\mu}},\qquad 0\le\xi\le1 .
\label{eq:fl}
\end{equation}
For the continuum model this law holds at all times~\cite{del2025generalized}. The derivation below uses that the walker is at the origin right after each reset. The time $\tau$ since the last reset has the density $re^{-r\tau}$ for $0<\tau<T$, and an atom of weight $e^{-rT}$ at $\tau=T$ that corresponds to trajectories with no reset in $[0,T]$.

We write $u=T-t_l=\tau(1-\xi)$ for the time elapsed since the last passage. Conditioning on $\tau$ gives
\begin{equation}
p(u|T)=\int_u^T d\tau\,r\,e^{-r\tau}\,\frac1\tau\,f_l\!\left(1-\frac u\tau\right)+\frac{e^{-rT}}{T}\,f_l\!\left(1-\frac uT\right).
\label{eq:tl_integral}
\end{equation}
With $\tau=u+v$, where $v=\tau\xi$ is the time between the last reset and the last passage, the integrand of the first term becomes $r\,\frac{\sin\pi\mu}{\pi}\,u^{-\mu}v^{\mu-1}e^{-r(u+v)}$. The integral over $v\in[0,T-u]$ is elementary, and with $\Gamma(\mu)\Gamma(1-\mu)=\pi/\sin\pi\mu$ we obtain
\begin{equation}
p(u|T)=\frac{r^{1-\mu}\,u^{-\mu}\,e^{-ru}}{\Gamma(1-\mu)}\;\frac{\gamma\bigl(\mu,\,r(T-u)\bigr)}{\Gamma(\mu)}+\frac{e^{-rT}}{T}\,f_l\!\left(1-\frac uT\right),\qquad 0\le u\le T,
\label{eq:tl_exact}
\end{equation}
where $\gamma(\mu,x)=\int_0^xt^{\mu-1}e^{-t}dt$ is the lower incomplete gamma function. The first term collects the trajectories with at least one reset, and the second term collects the trajectories with no reset. For $r\to0$ it reduces to the free law $f_l$. For $\alpha=0$ it gives the last-passage time of Brownian motion with resetting to its starting point~\cite{tazbierski2025last,wang2026pairwise}.

\subsection{Long-time Gamma law}
\label{subsec:tl_gamma}

For $rT\gg1$ the second term of Eq.~\eqref{eq:tl_exact} is exponentially small and $\gamma(\mu,r(T-u))\to\Gamma(\mu)$. The time $u=T-t_l$ elapsed since the last passage then follows a Gamma law with shape $1-\mu$ and rate $r$,
\begin{equation}
p(u)=\frac{r^{1-\mu}\,u^{-\mu}\,e^{-ru}}{\Gamma(1-\mu)},\qquad \bigl\langle e^{-ku}\bigr\rangle=\Bigl(1+\frac kr\Bigr)^{\mu-1},
\label{eq:tl_gamma}
\end{equation}
with mean $\langle u\rangle=(1-\mu)/r$ and variance $(1-\mu)/r^2$. The shape parameter $1-\mu$ carries the sluggishness. The divergence $u^{-\mu}$ at small $u$ is inherited from the edge $\xi\to1$ of the free law~\eqref{eq:fl}. For $\alpha=0$ the shape parameter is $1/2$, and as $\alpha$ grows the shape increases toward one, so the law approaches an exponential distribution. The result is an instance of a standard identity. Since $u=\tau(1-\xi)$, with $\tau$ exponential and $1-\xi$ an independent Beta variable of parameters $(1-\mu,\mu)$, the product of a Gamma variable of shape $1$ and a Beta$(1-\mu,\mu)$ variable is a Gamma variable of shape $1-\mu$.

The mean is exact at any $rT$. The time since the last reset has the mean $\langle\tau\rangle=(1-e^{-rT})/r$, which collects the density $re^{-r\tau}$ on $(0,T)$ and the atom at $\tau=T$, and the free law~\eqref{eq:fl} gives $\langle1-\xi\rangle=1-\mu$. Since $\xi$ and $\tau$ are independent,
\begin{equation}
\langle T-t_l\rangle=\frac{(1-\mu)\bigl(1-e^{-rT}\bigr)}{r},\qquad
\langle t_l\rangle=T-\frac{(1-\mu)\bigl(1-e^{-rT}\bigr)}{r}.
\label{eq:tl_mean}
\end{equation}
For $r\to0$ this gives $\langle t_l\rangle=\mu T$, the mean of the free law, and for $rT\gg1$ it gives $\langle T-t_l\rangle\to(1-\mu)/r$, the mean of the Gamma law.

\begin{figure}
\centering
\includegraphics[width=\linewidth]{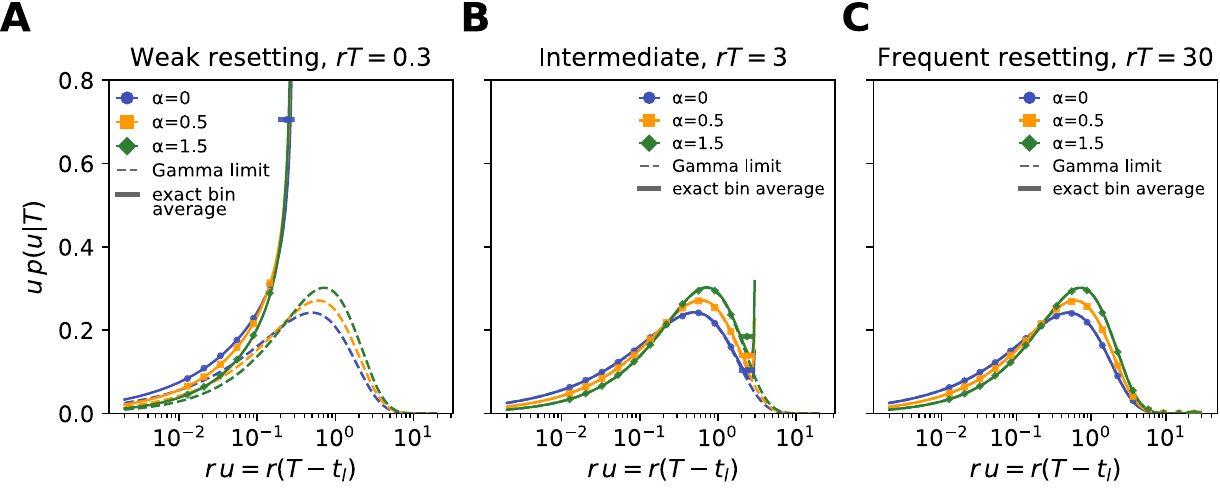}
\caption{Distribution of the time $u=T-t_l$ elapsed since the last passage through the origin, shown as $u\,p(u|T)$ (the density per logarithmic interval) as a function of $ru$, at weak ($rT=0.3$, \textbf{(A)}), intermediate ($rT=3$, \textbf{(B)}) and frequent ($rT=30$, \textbf{(C)}) resetting, for sluggishness parameters $\alpha=\{0, 0.5, 1.5\}$ (see legend). Symbols are averages over logarithmic bins of $N=10^6$ simulated trajectories, generated in 10 independent blocks, with error bars given by the standard error across blocks. In the last bin, where the density diverges at $u=T$, the horizontal bar shows the exact bin average. Solid lines are given by the exact law, Eq.~\eqref{eq:tl_exact}, and dashed lines by the long-time Gamma law, Eq.~\eqref{eq:tl_gamma}.}
\label{fig:lastpassage}
\end{figure}

At weak resetting the law of $u$ is broad (Fig.~\ref{fig:lastpassage}), confined to $u\le T$, and diverges at the end of its support, as the free law does. As $rT$ grows the law moves away from the free form. At intermediate resetting it already coincides with the Gamma law of Eq.~\eqref{eq:tl_gamma} except close to the end of the support, and at frequent resetting it is indistinguishable from it. The position and height of the maximum of $u\,p(u|T)$ move to larger $ru$ and rise as $\alpha$ increases, because the shape of the Gamma law grows toward that of an exponential.

\subsection{Exact vanishing of occupation--last-passage correlations}
\label{subsec:correlations}

The occupation-time distribution~\cite{frimpong2026} is not needed here, but reflection symmetry gives one exact relation among the global times. Under $x\mapsto-x$ the last-passage time is unchanged while $t_+\mapsto T-t_+$. The conditional distribution of $t_+$ at fixed $t_l$ is therefore symmetric about $T/2$, so $\langle t_+|t_l\rangle=T/2$ ($t_+$ is mean independent of $t_l$, which is weaker than statistical independence) and
\begin{equation}
\mathrm{Cov}(t_+,t_l)=0
\qquad\text{for all }\alpha,\ r\text{ and }T.
\label{eq:cov0}
\end{equation}
More generally $\mathrm{Cov}(t_+,g(t_l))=0$ for every function $g$ with finite variance, in particular for the powers $t_l^m$. For $\alpha=0$ this reproduces the absence of correlation between the occupation time and powers of the last-passage time that was found for Brownian motion with resetting~\cite{wang2026pairwise}. The argument uses only the reflection symmetry of the process and of the resetting protocol, and therefore holds equally for non-Poissonian reset times.

\section{Conclusion}
\label{sec:conclusions}

We have derived the extremal and global-time statistics of a sluggish random walk under Poissonian resetting to the origin. The nonlinear coordinate $Y=2\mu\,\mathrm{sgn}(x)|x|^{1/(2\mu)}$ converts the position-dependent diffusion into a signed Bessel process and isolates the effect of sluggishness in the Bessel index $\mu=1/(\alpha+2)$. Combined with renewal theory, this gives an exact Laplace-domain distribution of the maximum and an exact double-Laplace representation of the time at which the maximum occurs.

The maximum has two complementary long-time structures. On the typical scale, the centered variable $\sqrt r\,Y_{\max}-a_T$ converges to the standard Gumbel law. In physical space this gives $M_{\rm typ}\propto r^{-\mu}[\ln(rT)]^{2\mu}$ and a width proportional to $r^{-\mu}[\ln(rT)]^{2\mu-1}$. On the far-tail scale $Y=wT$, the rate function changes from $\sqrt r\,w$ to $r+w^2/4$ at $w_c=2\sqrt r$. It does not depend on $\alpha$ in the variable $Y$ and reduces at $\alpha=0$ to the Brownian result~\cite{hartmann2024distribution}. The transition of the rate function separates maxima produced by a single anomalously long interval of duration $Y/(2\sqrt r)<T$ that ends before the window does from maxima produced by an interval that occupies essentially the whole window.

The global times retain different signatures of renewal. The time of the maximum becomes uniform only in the bulk of the window. A right boundary layer of width $1/r$ and a broader left depletion associated with the finite climbing time generate the positive shift $\langle t_M\rangle-T/2\sim\ln(rT)/(4r)$. In contrast, the last passage localizes near the end of the window, with $T-t_l$ converging to a Gamma law of shape $1-\mu$. Reflection symmetry implies $\mathrm{Cov}(t_+,t_l)=0$ exactly.

The analysis relies on resetting to the origin and on exponential waiting times between resets. Natural extensions are resetting to a point away from the origin, non-Poissonian resetting, and extremal observables under distributed or partial resetting. For power-law waiting times the extreme-value statistics can leave the Gumbel class, as found for a ballistic process with random resets~\cite{villarroel2018continuous}, and the corresponding problem for sluggish walks is open.

\section*{Appendix}
\appendix
\section{Derivations}
\subsection{Matching condition at the origin}
\label{app:bessel}

For $x>0$ the two independent solutions of the backward equation of Sec.~\ref{subsec:backward} are $\Phi_0(z)=z^{\mu}I_{-\mu}(z)$ and $\Phi_1(z)=z^{\mu}I_{\mu}(z)$ with $z=\sqrt s\,y(x)$. Near the origin $\Phi_0(z)=c_\mu+O(z^2)$ and $\Phi_1(z)\propto z^{2\mu}\propto|x|$. Hence $\Phi_0(\sqrt s|y|)$ is even in $x$, $\mathrm{sgn}(x)\Phi_1(\sqrt s|y|)$ is odd, and both are continuously differentiable across $x=0$. The matching condition, continuity of $\psi$ and $\psi'$ at $x=0$, therefore selects the solutions $a\,\Phi_0+b\,\mathrm{sgn}(x)\Phi_1$ evaluated at $\sqrt s|y|$. The condition also follows from integrating $\psi''=s|x|^{\alpha}\psi$ across the origin, because the right-hand side is locally integrable.

For $x<0$ the solution is $a\,\Phi_0-b\,\Phi_1$. Using
\begin{equation}
z^{\mu}K_\mu(z)=\frac{\pi}{2\sin\pi\mu}\,\bigl[\Phi_0(z)-\Phi_1(z)\bigr],
\end{equation}
boundedness as $x\to-\infty$ requires $a=b$, since otherwise the exponentially growing part of $\Phi_0$ and $\Phi_1$ survives. The choice $a=b=1$ gives $\Phi_0+\Phi_1=g_\mu$ for $x\ge0$ and $\Phi_0-\Phi_1=(2\sin\pi\mu/\pi)\,z^{\mu}K_\mu(z)$ for $x<0$, which is $h_s$ of Eq.~\eqref{eq:hs}. Adding the particular solution $1/s$ of Eq.~\eqref{eq:survival_ode} and imposing the absorbing condition at $M$ gives Eq.~\eqref{eq:Q0_general}.

The matching condition is the symmetric, non-sticky extension of the process at the origin. Without resetting, the lattice walk of Sec.~\ref{subsec:model} is a symmetric nearest-neighbor chain with reversible weights proportional to $(|x|/b)^{\alpha}+2$. These converge to a measure proportional to $|x|^{\alpha}dx$ (to $dx$ for $\alpha=0$) with no point mass at the origin, because a single site carries negligible weight compared with any fixed neighborhood. The walker is therefore not sticky at the origin, and the left--right symmetry of the chain excludes a bias toward either side.

\subsection{Long-time analysis of the maximum}
\label{app:maxlong}

\subsubsection{Dominant pole}
\label{app:pole}

Write $u=s+r$ for the shifted Laplace variable and $G=G_\mu(M,u)$. Since $1-G=u\tilde Q_0(M,u)$, the denominator of Eq.~\eqref{eq:Qr} is $s+rG=u\,[1-r\tilde Q_0(M,u)]$, so apart from $u=0$ its zeros solve $r\tilde Q_0(M,u)=1$. Here $Q_0(M,t)=\mathrm{Prob}[M_t<M]$ is positive and non-increasing in $t$. For real $u>0$, $\tilde Q_0(M,u)$ decreases strictly from $+\infty$ as $u\to0^+$, because $Q_0\sim t^{-\mu}$ with $\mu<1$, to zero as $u\to\infty$. Since $\tilde Q_0(M,r)=[1-G_\mu(M,r)]/r<1/r$, the equation $r\tilde Q_0(M,u)=1$ has exactly one real root $u_0\in(0,r)$, which gives the pole $s_0=u_0-r\in(-r,0)$. For $u=a+ib$ with $a\ge0$ and $b>0$ the function $e^{-at}Q_0(M,t)$ is positive and non-increasing, so $\int_0^\infty e^{-at}Q_0\sin(bt)\,dt>0$ and $\mathrm{Im}\,\tilde Q_0<0$, which excludes complex roots. Because $G_\mu(M,u)$ is the Laplace transform of a probability density, it is analytic for $\mathrm{Re}\,u>0$. The other singularities of $\tilde Q_r$ are the branch point of $G_\mu$ at $u=0$ and the zeros of $g_\mu(\sqrt u\,Y)$ on the negative real $u$ axis, and all of them lie at $\mathrm{Re}\,s\le-r$. This establishes the existence and location of the real pole $s_0$ and the absence of competing poles in $\mathrm{Re}\,s>-r$. We take the remaining contribution after deformation of the Bromwich contour to decay as $e^{-rT}$ times a power of $T$. This is an asymptotic estimate, not a uniform bound. We use it for a level $M=M(T)$ in the extreme-value scaling regime, where $\lambda(M)T=O(1)$, and the comparison with simulations in Fig.~\ref{fig:gumbel} tests it.

\subsubsection{Large-deviation bounds}
\label{app:ld}

\textit{Uniform bounds for the free first passage.} In the variable $y$, $|y_t|=\beta_{2t}$ in law, where $\beta$ is a standard Bessel process of dimension $d\in[1,2)$ started at the origin, and $\sigma_Y$ is the first time at which $y$ reaches $Y$. The squared Bessel processes of dimensions $1\le d\le2$ obey stochastic differential equations with the same noise coefficient and ordered constant drifts, so by the comparison theorem for one-dimensional diffusions (Ref.~\cite{revuz2013continuous}, Chs.~IX and~XI) the maximum of $\beta$ over $[0,t]$ is stochastically bounded above by the maximum of $|\mathbf W_s|$ for a planar Brownian motion $\mathbf W$ and below by the maximum of $|W_s|$ for a one-dimensional Brownian motion $W$. Both bounds are exact in law at $d=2$ and $d=1$, respectively.

For the upper bound, $e^{\lambda|\mathbf W_s|}$ is a nonnegative submartingale for every $\lambda>0$, because it is a convex increasing function of the submartingale $|\mathbf W_s|$. Doob's maximal inequality then gives $\mathrm{Prob}[\max_{s\le t}|\mathbf W_s|\ge Y]\le e^{-\lambda Y}\,\mathbb E\,e^{\lambda|\mathbf W_t|}$. The Rayleigh density of $|\mathbf W_t|$ gives $\mathbb E\,e^{\lambda|\mathbf W_t|}\le\sqrt{2\pi}\,(1+\lambda\sqrt t)\,e^{\lambda^2t/2}$, and the choice $\lambda=Y/t$ with $t=2\tau$ yields
\[
\mathrm{Prob}[\sigma_Y\le\tau]\le\sqrt{2\pi}\Bigl(1+\frac{Y}{\sqrt{2\tau}}\Bigr)e^{-Y^2/4\tau}\le C_\varepsilon\,e^{-(1-\varepsilon)Y^2/4\tau},
\]
uniformly in $Y$ and $\tau$. For the lower bound, the walker reaches $+Y$ before $\tau$ with at least half the probability that $|y|$ reaches $Y$, by the symmetry $y\to-y$. Therefore $\mathrm{Prob}[\sigma_Y^{+}\le\tau]\ge\tfrac12\mathrm{Prob}[\max_{s\le2\tau}|W_s|\ge Y]\ge\mathrm{Prob}[W_{2\tau}\ge Y]=\tfrac12\mathrm{erfc}\bigl(Y/2\sqrt\tau\bigr)$, and the bound $\mathrm{erfc}(x)\ge(2/\sqrt\pi)\,e^{-x^2}/(x+\sqrt{x^2+2})$ gives the left inequality of Eq.~\eqref{eq:fpt_bounds}. These bounds apply to the first passage to $+Y$ as well, which is the one that enters the maximum.

\textit{Upper bound.} Expanding $\tilde F_r/s=(r+s)G_\mu(M,r+s)/[s(s+rG_\mu)]$ to first order in $G_\mu$ gives $(r+s)G_\mu(M,r+s)/s^2$. Since $G_\mu(M,r+s)$ is the Laplace transform of $h_Y(\tau)=e^{-r\tau}F_0(M,\tau)$, the inverse of this first-order term is the integral in Eq.~\eqref{eq:union}. It is the mean number $\langle N\rangle$ of intervals whose maximum exceeds $M$, and the union bound gives $\mathrm{Prob}[M_T>M]\le\langle N\rangle$. Write $H(\tau)=\mathrm{Prob}[\sigma_Y^{+}\le\tau]$, so that $F_0(M,\tau)\,d\tau=dH$, and $w(\tau)=[1+r(T-\tau)]e^{-r\tau}$. Integrating by parts, $\langle N\rangle=w(T)H(T)-\int_0^TH(\tau)\,w'(\tau)\,d\tau$, with $0\le-w'(\tau)\le r(2+rT)\,e^{-r\tau}$. The upper bound of Eq.~\eqref{eq:fpt_bounds} then gives
\[
\langle N\rangle\le C_\varepsilon\bigl[1+rT(2+rT)\bigr]\max_{0<\tau\le T}\exp\Bigl[-r\tau-(1-\varepsilon)\frac{Y^2}{4\tau}\Bigr].
\]
For $Y=wT$ the maximum equals $e^{-T\Phi(\sqrt{1-\varepsilon}\,w)}$ with $\Phi$ the function of Eq.~\eqref{eq:ldp}, since $\min_{0<v\le1}[rv+w^2/(4v)]=\Phi(w)$. The polynomial factor does not change the exponential rate, and $\varepsilon\to0$ with the continuity of $\Phi$ gives $\limsup_{T\to\infty}T^{-1}\ln\mathrm{Prob}[M_T>M]\le-\Phi(w)$.

\textit{Lower bound.} The initial interval alone reaches $M$ before its reset and before $T$ with probability $H_T=\int_0^Te^{-r\tau}F_0(M,\tau)\,d\tau=\mathbb E[e^{-r\sigma_Y^{+}};\sigma_Y^{+}\le T]$, so $\mathrm{Prob}[M_T>M]\ge H_T\ge e^{-r\tau_1}H(\tau_1)$ for every $\tau_1\le T$. With $\tau_1=\min(\tau^*,T)$, $\tau^*=Y/(2\sqrt r)=wT/(2\sqrt r)$, and the lower bound of Eq.~\eqref{eq:fpt_bounds}, the exponent is $r\tau_1+Y^2/(4\tau_1)$, up to a polynomial factor. This equals $\sqrt r\,w$ for $w<2\sqrt r$, where $\tau^*<T$, and $r+w^2/4$ for $w>2\sqrt r$, where $\tau_1=T$ and the event is that of no reset, which has probability $e^{-rT}$. Hence $\liminf_{T\to\infty}T^{-1}\ln\mathrm{Prob}[M_T>M]\ge-\Phi(w)$, which together with the upper bound proves Eq.~\eqref{eq:ldp}.

\subsection{Derivations for the time of the maximum}
\label{app:tmax}

\subsubsection{Path decomposition}
\label{app:decomp}

We derive Eq.~\eqref{eq:joint} from the strong Markov property, which applies because the position together with the exponential reset clock is a Markov process. The argmax $t_M$ is almost surely unique, so the event $\{t_M\in dt_1\}$ is well defined. Let $\sigma_M$ be the first time at which the path reaches $M$, with density $F_r(M,t)$. On the event $\{M_T\in[M,M+dM]\}$ the path stays below $M+dM$ and leaves the strip of width $dM$ below this barrier within a time of order $dM^2/D(M)=M^{\alpha}\,dM^2$, which vanishes with $dM$ at fixed $M$, so $t_M-\sigma_M\to0$ as $dM\to0$. Up to a vanishing error, $\{M_T\in dM,\ t_M\in dt_1\}$ is therefore the event that $\sigma_M\in dt_1$ and that the path restarted at $M$ stays below $M+dM$ during the remaining time $t_2=T-t_1$. Let $Q_r(x_0,t\,|\,B)$ be the survival probability of the reset process started at $x_0$ with an absorbing barrier at $B$. Since $Q_r(B,t\,|\,B)=0$ for every $B$, differentiating with respect to $B$ gives $\partial_{x_0}Q_r+\partial_BQ_r=0$ at $x_0=B$, and hence $Q_r(M,t_2\,|\,M+dM)=q_r(M,t_2)\,dM+o(dM)$ with $q_r=-\partial_{x_0}Q_r|_{x_0=M}$. The probability of the event is $F_r(M,t_1)\,q_r(M,t_2)\,dt_1\,dM$, which is Eq.~\eqref{eq:joint}. As a check, Eqs.~\eqref{eq:Qr}, \eqref{eq:Fr} and~\eqref{eq:qr} with $\partial_YG_\mu=-\sqrt u\,\eta\,G_\mu$ give the exact identity $\tilde F_r(M,s)\,\tilde q_r(M,s)=\partial_M\tilde Q_r(M,s)$, so setting $s_1=s_2$ in Eq.~\eqref{eq:Phat} integrates the joint law over $t_M$ to the exact density of the maximum.

\subsubsection{Mean time of the maximum}
\label{app:mean}

We write Eq.~\eqref{eq:ptilde} as $\hat p(s_1,s_2)=\int_0^\infty dY\,\tilde F(s_1,Y)\,K(s_2,Y)$ with
\begin{gather}
    \tilde F=\frac{u\,G}{D},\qquad K=\frac{\sqrt u\,\eta}{D},\qquad u=s+r,\quad Z=\sqrt u\,Y, \notag\\[4pt] G=\frac{c_\mu}{g_\mu(Z)},\quad \eta=\frac{g_\mu'(Z)}{g_\mu(Z)},\quad D=s+rG .
\end{gather}
By Eq.~\eqref{eq:tM_mean} we need $\tfrac12\int dY\,[\tilde F\,\partial_sK-K\,\partial_s\tilde F]$ at fixed $Y$. Since $\partial_sZ=Z/(2u)$ and $dG/dZ=-G\eta$, one has $\partial_sG=-G\theta$ with $\theta=\eta Z/(2u)$, which gives
\begin{equation}
\partial_s\tilde F=\frac{G(1-u\theta)}{D}-\frac{uG(1-rG\theta)}{D^2},\qquad
\partial_sK=\frac{\eta+Z\eta'}{2\sqrt u\,D}-\frac{\sqrt u\,\eta\,(1-rG\theta)}{D^2}.
\end{equation}
The terms proportional to $1-rG\theta$ cancel in $\tilde F\,\partial_sK-K\,\partial_s\tilde F$. Using $u\theta=\eta Z/2$, what remains is
\begin{equation}
\tilde F\,\partial_sK-K\,\partial_s\tilde F=\frac{\sqrt u\,G}{2D^2}\Bigl[-\eta+Z\bigl(\eta'+\eta^2\bigr)\Bigr].
\end{equation}
The differential equation of $g_\mu$ (Sec.~\ref{subsec:backward}) gives $\eta'+\eta^2=1-(1-2\mu)\eta/Z$, so the bracket equals $Z-d\,\eta$ with $d=2(1-\mu)$. With $dY=dZ/\sqrt u$ the dependence on $u$ drops out, and Eq.~\eqref{eq:tM_exact} follows.

Because $G'=-G\eta$, the term proportional to $d$ is a total derivative. Using $G(0)=1$ and $G(\infty)=0$,
\begin{equation}
\int_0^\infty dZ\,\frac{G'}{(s+rG)^2}=\Bigl[-\frac{1}{r(s+rG)}\Bigr]_0^\infty=-\frac{1}{s(s+r)} ,
\end{equation}
so Eq.~\eqref{eq:tM_exact} equals $\tfrac14\bigl[I(s)-d/(s(s+r))\bigr]$ with $I(s)=\int_0^\infty Z\,G\,(s+rG)^{-2}\,dZ$. For $\alpha=0$ one finds $I(s)=\ln(1+r/s)/(rs)$, which gives Eq.~\eqref{eq:tM_alpha0}.

For small $s$ we change variable to $\lambda=rG(Z)$, so that $dZ=-d\lambda/(\lambda\eta)$ and $I(s)=r^{-1}\int_0^r d\lambda\,Z(\lambda)/[\eta\,(s+\lambda)^2]$. The integral is dominated by $\lambda\sim s\ll r$, where $Z$ is large. There $\eta=1-(\tfrac12-\mu)/Z+O(Z^{-2})$, so $Z/\eta=Z+(\tfrac12-\mu)+O(1/Z)$, and Eq.~\eqref{eq:G_asym} gives $Z(\lambda)=\ln(A_\mu r/\lambda)+(\tfrac12-\mu)\ln Z(\lambda)+o(1)$. The weight $s/(s+\lambda)^2$ is normalized on $(0,\infty)$ and averages $\ln(s/\lambda)$ to zero. Therefore $r\,s\,I(s)=Z_s+(\tfrac12-\mu)+o(1)$, where $Z_s$ solves $A_\mu Z_s^{1/2-\mu}e^{-Z_s}=s/r$. Collecting terms,
\begin{equation}
4\,r\,s\int_0^\infty dT\,e^{-sT}\Bigl[\langle t_M\rangle_T-\frac T2\Bigr]=Z_s+\mu-\tfrac32+o(1).
\end{equation}
Finally, $Z_s$ is the Laplace-domain counterpart of $a_T+\gamma_E$, because $\int_0^\infty e^{-sT}\ln(rT)\,dT=[\ln(r/s)-\gamma_E]/s$. Equation~\eqref{eq:tM_shift} follows.

\subsubsection{Boundary layers}
\label{app:layers}

\textit{Right layer.} With $\lambda=rG_\mu(M,r)$ one has $d\lambda=-\sqrt r\,\eta\,\lambda\,dY$, where $\eta$ is evaluated at $\sqrt r\,Y$ and can be regarded as a function of $\lambda$. Hence $\int dY\,\lambda e^{-\lambda T}=r^{-1/2}\int_0^r e^{-\lambda T}\,\eta(\lambda)^{-1}\,d\lambda\simeq1/(\sqrt r\,T)$, because the integral is controlled by $\lambda\sim1/T$. This corresponds to large $Y$, $\zeta\simeq\ln(rT)$, where $G\ll1$ and $\eta\to1$. Multiplying by the inverse transform $k(t)=e^{-rt}/\sqrt{\pi t}+\sqrt r\,\mathrm{erf}\sqrt{rt}$ of the second factor gives $T\,p=k(t)/\sqrt r=\kappa(rt)$ with $t=T-t_M$. The factors of $\eta$, which tend to one at large $Y$, produce the relative correction $(\tfrac12-\mu)/a_T$ at small $t$.

\textit{Left layer.} The Laplace transform of $h_Y(t)=e^{-rt}F_0(M,t)$ is $G_\mu(M,r+s)=G_\mu(M,r)\,\tilde\pi_Y(s)$, where $\tilde\pi_Y$ is the Laplace transform of the climbing-time density $\pi_Y$. Hence $h_Y=G_\mu\,\pi_Y$ and $r\int_0^th_Y=\lambda\,\Pi_Y$, so that $F_r=\lambda\,[\Pi_Y+\pi_Y/r]$. The remaining factor $\sqrt r\,\eta\,\lambda\,e^{-\lambda T}dY$ equals $T^{-1}$ times the density of the level of the global maximum, because $\partial_Y e^{-rTG_\mu}=T\sqrt r\,\eta\,\lambda\,e^{-\lambda T}$. This gives Eq.~\eqref{eq:leftlayer}. The mean of $\pi_Y$ follows from $\tilde\pi_Y(s)=1-s\tau_Y+\dots$.

The integrated deviations from the uniform bulk are $\int_0^\infty(1-\Pi_Y-\pi_Y/r)\,dt=\tau_Y-1/r$ on the left and $\int_0^\infty(\kappa(\rho)-1)\,d\rho=\int_0^\infty e^{-\rho}/\sqrt{\pi \rho}\,d\rho-\int_0^\infty\mathrm{erfc}\sqrt\rho\,d\rho=1-\tfrac12=\tfrac12$ on the right, in units of $1/r$, which are the weights used in Sec.~\ref{subsec:tmax_mean}.

\section{Numerical implementation}
\label{app:numerics}

\subsection{Stochastic simulation}

Trajectories were generated from the time-change representation $x(t)=B(\tau(t))$ with $dt=|B|^{\alpha}d\tau$, accumulated by the trapezoidal rule. An interval of duration $e$ was advanced with the operational-time step $\Delta\tau=\Delta\tau_0\,e^{2\mu}$, $\Delta\tau_0=2\times10^{-3}$. In each step the maximum was drawn from the Brownian-bridge law given the endpoints~\cite{karatzas1991brownian} and its time was placed at the midpoint of the step. The last zero was placed by linear interpolation between the endpoints when they have opposite signs. For same-sign endpoints the bridge touches zero with probability $\exp(-B_nB_{n+1}/\Delta\tau)$, where $B_n$ and $B_{n+1}$ are the values of the Brownian path at the two ends of the step, and the touch was placed at the same interpolated point, a heuristic accurate to one step.

For $\alpha=0$ the maximum and its time were sampled exactly from the arcsine law for the time and, given the time $\theta$ of the maximum, the density proportional to $m\,e^{-m^2/4\theta}$ for the value. For $\alpha>0$ and large $rT$, complete intervals were resampled from $R=10$ independent pools of $2\times10^{5}$ simulated intervals, each stored with its maximum, the time of its maximum and its last zero. The truncated last interval was always simulated afresh. All simulated curves use $N=10^6$ trajectories in $R=10$ blocks, and error bars are the standard error across blocks, which includes the dependence among trajectories drawn from the same pool.

\subsection{Laplace inversion}

Laplace transforms were inverted with the fixed-Talbot rule~\cite{talbot1979accurate,abate2006unified}. The density of $t_M$ was obtained by quadrature over $Y$ after separate inversions of the two factors of Eq.~\eqref{eq:ptilde}. Tail probabilities were computed from the exceedance transform $\int_0^\infty e^{-sT}\,\mathrm{Prob}[M_T\ge M]\,dT=(s+r)\,G_\mu(M,s+r)\big/\{s\,[s+r\,G_\mu(M,s+r)]\}$, which avoids the cancellation in $1-\mathrm{Prob}[M_T<M]$, in extended precision with $\lfloor T\Phi(w)/2.3\rfloor+40$ digits.

\subsection{Validation}

Halving the step ($\Delta\tau_0=4\times10^{-3}$, $2\times10^{-3}$, $10^{-3}$ at $rT=3$) changed the distribution function of $M_T$ and the means of $t_M$ and $T-t_l$ by amounts comparable to their statistical errors. For $\alpha=0$ the stepped algorithm agrees with the exact sampler. For $\alpha>0$ the pool method agrees with direct simulation of complete trajectories at $rT=8$ and $rT=10^3$. The pooled mean of $t_M$ depends on the pool size, because it is set by the longest intervals. At $rT=10^3$, pools of $P=2\times10^{3}$ and $P=2\times10^{4}$ intervals bias $\langle t_M\rangle-T/2$ low by $42/r$ and $4.2/r$, a bias that falls roughly as $1/P$, so the pooled mean offset is shown only up to $rT=300$. Pooled far-tail probabilities are also unreliable, since a pool holds only of order one interval that reaches the highest levels, and the far tail is taken from the exact inversion. The simulations therefore test the exact transforms directly, up to $rT=300$ for the mean offset of $t_M$ and up to $rT=10^3$ for the other quantities, whereas the asymptotic laws at larger $rT$ are compared with the numerical inversion of the exact transforms.

%





\bibliography{references}

@article{majumdar2020extreme,
  title={Extreme value statistics of correlated random variables: a pedagogical review},
  author={Majumdar, Satya N and Pal, Arnab and Schehr, Gr{\'e}gory},
  journal={Physics Reports},
  volume={840},
  pages={1--32},
  year={2020},
  publisher={Elsevier}
}

@book{majumdar2024statistics,
  title={Statistics of Extremes and Records in Random Sequences},
  author={Majumdar, Satya N and Schehr, Gr{\'e}gory},
  year={2024},
  publisher={Oxford University Press}
}

@article{levy1940certains,
  title={Sur certains processus stochastiques homog{\`e}nes},
  author={L{\'e}vy, Paul},
  journal={Compositio mathematica},
  volume={7},
  pages={283--339},
  year={1940}
}

@book{feller1991introduction,
  title={An introduction to probability theory and its applications},
  author={Feller, William},
  volume={2},
  year={1991},
  publisher={John Wiley \& Sons}
}

@article{del2025generalized,
  title={Generalized arcsine laws for a sluggish random walker with subdiffusive growth},
  author={{Del Vecchio Del Vecchio}, Giuseppe and Majumdar, Satya N},
  journal={Journal of Statistical Mechanics: Theory and Experiment},
  volume={2025},
  number={2},
  pages={023207},
  year={2025},
  publisher={IOP Publishing}
}

@article{hartmann2025exact,
  title={Exact joint distributions of three global characteristic times for {Brownian} motion},
  author={Hartmann, Alexander K and Majumdar, Satya N},
  journal={Physical Review E},
  volume={111},
  number={4},
  pages={044134},
  year={2025},
  publisher={APS}
}

@article{evans2011diffusion,
  title={Diffusion with stochastic resetting},
  author={Evans, Martin R and Majumdar, Satya N},
  journal={Physical Review Letters},
  volume={106},
  number={16},
  pages={160601},
  year={2011},
  publisher={APS}
}

@article{evans2020stochastic,
  title={Stochastic resetting and applications},
  author={Evans, Martin R and Majumdar, Satya N and Schehr, Gr{\'e}gory},
  journal={Journal of Physics A: Mathematical and Theoretical},
  volume={53},
  number={19},
  pages={193001},
  year={2020},
  publisher={IOP Publishing}
}

@article{singh2021extremal,
  title={Extremal statistics for stochastic resetting systems},
  author={Singh, Prashant and Pal, Arnab},
  journal={Physical Review E},
  volume={103},
  number={5},
  pages={052119},
  year={2021},
  publisher={APS}
}

@article{wang2026pairwise,
  title={Pairwise correlations of global times in one-dimensional {Brownian} motion under stochastic resetting},
  author={Wang, Yihao and Chen, Hanshuang},
  journal={Physical Review E},
  volume={113},
  number={1},
  pages={014131},
  year={2026},
  publisher={APS}
}

@article{zodage2023sluggish,
  title={A sluggish random walk with subdiffusive spread},
  author={Zodage, Aniket and Allen, Rosalind J and Evans, Martin R and Majumdar, Satya N},
  journal={Journal of Statistical Mechanics: Theory and Experiment},
  volume={2023},
  number={3},
  pages={033211},
  year={2023},
  publisher={IOP Publishing}
}

@misc{frimpong2026,
  title={Sluggish random walks with resetting: nonequilibrium aspects},
  author={Frimpong, Theophilus Dwamena and Bresque, L\'ea and Rold\'an, \'Edgar},
  year={2026},
  eprint={2610.21962},
  archivePrefix={arXiv},
  primaryClass={cond-mat.stat-mech}
}

@article{boyer2026ultraslow,
  title={Ultra slow sub-logarithmic diffusion of a sluggish random walker subject to resetting with memory},
  author={Boyer, Denis and Majumdar, Satya N},
  journal={Journal of Statistical Mechanics: Theory and Experiment},
  volume={2026},
  pages={063201},
  year={2026},
  publisher={IOP Publishing}
}

@book{revuz2013continuous,
  title={Continuous martingales and {Brownian} motion},
  author={Revuz, Daniel and Yor, Marc},
  year={2013},
  publisher={Springer Science \& Business Media}
}

@book{borodin2012handbook,
  title={Handbook of {Brownian} motion-facts and formulae},
  author={Borodin, Andrei N and Salminen, Paavo},
  year={2012},
  publisher={Birkh{\"a}user}
}

@book{olver2010nist,
  editor    = {Olver, Frank W. J. and Lozier, Daniel W. and Boisvert, Ronald F. and Clark, Charles W.},
  title     = {{NIST} Handbook of Mathematical Functions},
  publisher = {Cambridge University Press},
  year      = {2010}
}

@book{redner2001guide,
  title={A guide to first-passage processes},
  author={Redner, Sidney},
  year={2001},
  publisher={Cambridge university press}
}

@book{gumbel2004statistics,
  title={Statistics of extremes},
  author={Gumbel, Emil Julius},
  year={2004},
  publisher={Courier Corporation}
}

@article{majumdar2015dynamical,
  title={Dynamical transition in the temporal relaxation of stochastic processes under resetting},
  author={Majumdar, Satya N and Sabhapandit, Sanjib and Schehr, Gr{\'e}gory},
  journal={Physical Review E},
  volume={91},
  number={5},
  pages={052131},
  year={2015},
  publisher={APS}
}

@article{majumdar2008time,
  title={On the time to reach maximum for a variety of constrained {Brownian} motions},
  author={Majumdar, Satya N and Randon-Furling, Julien and Kearney, Michael J and Yor, Marc},
  journal={Journal of Physics A: Mathematical and Theoretical},
  volume={41},
  number={36},
  pages={365005},
  year={2008}
}

@book{karatzas1991brownian,
  author    = {Karatzas, Ioannis and Shreve, Steven E.},
  title     = {Brownian Motion and Stochastic Calculus},
  edition   = {2},
  publisher = {Springer},
  address   = {New York},
  year      = {1991}
}

@article{talbot1979accurate,
  title={The accurate numerical inversion of {Laplace} transforms},
  author={Talbot, Alan},
  journal={IMA Journal of Applied Mathematics},
  volume={23},
  number={1},
  pages={97--120},
  year={1979},
  publisher={Oxford University Press}
}

@article{abate2006unified,
  title={A unified framework for numerically inverting {Laplace} transforms},
  author={Abate, Joseph and Whitt, Ward},
  journal={INFORMS Journal on Computing},
  volume={18},
  number={4},
  pages={408--421},
  year={2006},
  publisher={INFORMS}
}

@book{morters2010brownian,
  title={Brownian Motion},
  author={M{\"o}rters, Peter and Peres, Yuval},
  year={2010},
  publisher={Cambridge University Press},
  address={Cambridge, UK}
}

@incollection{majumdar2005brownian,
  title={Brownian functionals in physics and computer science},
  author={Majumdar, Satya N},
  booktitle={The legacy of Albert Einstein: A collection of essays in celebration of the year of physics},
  pages={93--129},
  year={2007},
  publisher={World Scientific}
}

@article{fortin2015applications,
  title={Applications of extreme value statistics in physics},
  author={Fortin, Jean-Yves and Clusel, Maxime},
  journal={Journal of Physics A: Mathematical and Theoretical},
  volume={48},
  number={18},
  pages={183001},
  year={2015},
  publisher={IOP Publishing}
}

@article{fisher1928limiting,
  title={Limiting forms of the frequency distribution of the largest or smallest member of a sample},
  author={Fisher, Ronald A and Tippett, Leonard H C},
  journal={Mathematical Proceedings of the Cambridge Philosophical Society},
  volume={24},
  pages={180},
  year={1928}
}

@article{gnedenko1943distribution,
  title={Sur la distribution limite du terme maximum d'une s{\'e}rie al{\'e}atoire},
  author={Gnedenko, Boris},
  journal={Annals of Mathematics},
  volume={44},
  pages={423},
  year={1943}
}

@article{mori2022time,
  title={Time to reach the maximum for a stationary stochastic process},
  author={Mori, Francesco and Majumdar, Satya N and Schehr, Gr{\'e}gory},
  journal={Physical Review E},
  volume={106},
  pages={054110},
  year={2022},
  doi={10.1103/PhysRevE.106.054110},
  eprint={2207.12329},
  archivePrefix={arXiv}
}

@article{sadhu2018generalized,
  title={Generalized arcsine laws for fractional {Brownian} motion},
  author={Sadhu, Tridib and Delorme, Mathieu and Wiese, Kay J{\"o}rg},
  journal={Physical Review Letters},
  volume={120},
  pages={040603},
  year={2018}
}

@article{singh2019generalised,
  title={Generalised `arcsine' laws for run-and-tumble particle in one dimension},
  author={Singh, Prashant and Kundu, Anupam},
  journal={Journal of Statistical Mechanics: Theory and Experiment},
  volume={2019},
  pages={083205},
  year={2019}
}

@article{bel2005weak,
  title={Weak ergodicity breaking in the continuous-time random walk},
  author={Bel, Golan and Barkai, Eli},
  journal={Physical Review Letters},
  volume={94},
  pages={240602},
  year={2005}
}

@article{singh2022extreme,
  title={Extreme value statistics and arcsine laws for heterogeneous diffusion processes},
  author={Singh, Prashant},
  journal={Physical Review E},
  volume={105},
  pages={024113},
  year={2022}
}

@article{gupta2022brief,
  title={Stochastic resetting: A (very) brief review},
  author={Gupta, Shamik and Jayannavar, Arun M},
  journal={Frontiers in Physics},
  volume={10},
  pages={789097},
  year={2022}
}

@article{reuveni2016optimal,
  title={Optimal stochastic restart renders fluctuations in first passage times universal},
  author={Reuveni, Shlomi},
  journal={Physical Review Letters},
  volume={116},
  pages={170601},
  year={2016}
}

@article{majumdar2021mean,
  title={Mean perimeter and area of the convex hull of a planar {Brownian} motion in the presence of resetting},
  author={Majumdar, Satya N and Mori, Francesco and Schawe, Hendrik and Schehr, Gr{\'e}gory},
  journal={Physical Review E},
  volume={103},
  pages={022135},
  year={2021}
}

@article{villarroel2018continuous,
  title={Continuous-time ballistic process with random resets},
  author={Villarroel, Javier and Montero, Miquel},
  journal={Journal of Statistical Mechanics: Theory and Experiment},
  volume={2018},
  pages={123204},
  year={2018}
}

@article{tazbierski2025last,
  title={Arcsine laws for {Brownian} motion with {Poissonian} resetting},
  author={Ta\'zbierski, Kacper and Magdziarz, Marcin},
  journal={Chaos},
  volume={35},
  pages={023163},
  year={2025},
  doi={10.1063/5.0253282},
  eprint={2502.03889},
  archivePrefix={arXiv}
}

@article{fa2003power,
  title={Power law diffusion coefficient and anomalous diffusion: Analysis of solutions and first passage time},
  author={Fa, Kwok Sau and Lenzi, Ervin K},
  journal={Physical Review E},
  volume={67},
  number={6},
  pages={061105},
  year={2003},
  publisher={APS}
}

@article{cherstvy2013anomalous,
  title={Anomalous diffusion and ergodicity breaking in heterogeneous diffusion processes},
  author={Cherstvy, Andrey G and Chechkin, Aleksei V and Metzler, Ralf},
  journal={New Journal of Physics},
  volume={15},
  pages={083039},
  year={2013}
}

@article{leibovich2019infinite,
  title={Infinite ergodic theory for heterogeneous diffusion processes},
  author={Leibovich, Nava and Barkai, Eli},
  journal={Physical Review E},
  volume={99},
  pages={042138},
  year={2019}
}

@article{stella2023universal,
  title={Universal singularities of anomalous diffusion in the {Richardson} class},
  author={Stella, Attilio L and Chechkin, Aleksei and Teza, Gianmaria},
  journal={Physical Review E},
  volume={107},
  pages={054118},
  year={2023}
}

@article{ray2020space,
  title={Space-dependent diffusion with stochastic resetting: A first-passage study},
  author={Ray, Somrita},
  journal={Journal of Chemical Physics},
  volume={153},
  pages={234904},
  year={2020}
}

@article{lenzi2022transient,
  title={Transient anomalous diffusion in heterogeneous media with stochastic resetting},
  author={Lenzi, Marcelo K and Lenzi, Ervin K and Guilherme, Luiz M S and Evangelista, Luiz R and Ribeiro, Haroldo V},
  journal={Physica A},
  volume={588},
  pages={126560},
  year={2022}
}

@article{menon2024random,
  title={Random search with resetting in heterogeneous environments},
  author={Menon, Leonardo and Anteneodo, Celia},
  journal={Physical Review E},
  volume={110},
  pages={054111},
  year={2024}
}

@article{ray2022expediting,
  title={Expediting {Feller} process with stochastic resetting},
  author={Ray, Somrita},
  journal={Physical Review E},
  volume={106},
  pages={034133},
  year={2022}
}

@article{delvecchio2026proxitaxis,
  title={Proxitaxis: an adaptive search strategy based on proximity and stochastic resetting},
  author={{Del Vecchio Del Vecchio}, Giuseppe and Kulkarni, Manas and Majumdar, Satya N and Sabhapandit, Sanjib},
  journal={Physical Review E},
  volume={113},
  pages={L042101},
  year={2026}
}

@article{majumdar2004exact,
  title={Exact maximal height distribution of fluctuating interfaces},
  author={Majumdar, Satya N and Comtet, Alain},
  journal={Physical Review Letters},
  volume={92},
  pages={225501},
  year={2004}
}

@article{majumdar2005airy,
  title={Airy distribution function: from the area under a {Brownian} excursion to the maximal height of fluctuating interfaces},
  author={Majumdar, Satya N and Comtet, Alain},
  journal={Journal of Statistical Physics},
  volume={119},
  pages={777},
  year={2005}
}

@article{randon2007distribution,
  title={Distribution of the time at which the deviation of a {Brownian} motion is maximum before its first-passage time},
  author={Randon-Furling, Julien and Majumdar, Satya N},
  journal={Journal of Statistical Mechanics: Theory and Experiment},
  volume={2007},
  pages={P10008},
  year={2007}
}

@article{schehr2010extreme,
  title={Extreme value statistics from the real space renormalization group: {Brownian} motion, {Bessel} processes and continuous time random walks},
  author={Schehr, Gr{\'e}gory and Le Doussal, Pierre},
  journal={Journal of Statistical Mechanics: Theory and Experiment},
  volume={2010},
  pages={P01009},
  year={2010}
}

@article{bray2013persistence,
  title={Persistence and first-passage properties in nonequilibrium systems},
  author={Bray, Alan J and Majumdar, Satya N and Schehr, Gr{\'e}gory},
  journal={Advances in Physics},
  volume={62},
  pages={225},
  year={2013}
}

@article{touchette2009large,
  title={The large deviation approach to statistical mechanics},
  author={Touchette, Hugo},
  journal={Physics Reports},
  volume={478},
  pages={1},
  year={2009}
}

@article{meylahn2015large,
  title={Large deviations for {Markov} processes with resetting},
  author={Meylahn, Janusz M and Sabhapandit, Sanjib and Touchette, Hugo},
  journal={Physical Review E},
  volume={92},
  pages={062148},
  year={2015}
}

@article{debicki2026distributional,
  title={Distributional and Extremal Behaviour of Brownian Motion with Exponential Resetting},
  author={Debicki, Krzysztof and Hashorva, Enkelejd and Michna, Zbigniew},
  journal={Methodology and Computing in Applied Probability},
  volume={28},
  number={3},
  pages={79},
  year={2026},
  publisher={Springer},
  doi={10.1007/s11009-026-10316-4}
}

@incollection{hartmann2024distribution,
  author={Hartmann, Alexander K. and Majumdar, Satya N. and Schehr, Gr{\'e}gory},
  title={The Distribution of the Maximum of Independent Resetting {Brownian} Motions},
  booktitle={Target Search Problems},
  publisher={Springer Nature Switzerland},
  address={Cham},
  pages={357--389},
  year={2024},
  note={arXiv:2309.17432, doi:10.1007/978-3-031-67802-8\_15}
}

\end{document}